\documentclass[sigconf,10pt]{acmart}

\renewcommand\footnotetextcopyrightpermission[1]{} % removes footnote with conference information in first column

\usepackage {color}
\usepackage{siunitx}
\usepackage{textcomp}
\usepackage{multirow}
\usepackage{graphicx}
\usepackage{geometry}
\usepackage{subfigure}
\usepackage{float}
\usepackage{url}
\usepackage{booktabs}
\hypersetup{
    colorlinks=true,
    linkcolor=red,
    urlcolor=magenta,
    citecolor=green,
}
\usepackage{siunitx}
\usepackage{bbding}
\usepackage{gensymb}
\usepackage{pifont}
\usepackage{arydshln}
\usepackage{diagbox}
\usepackage{gensymb}
\usepackage{fontawesome}
\usepackage{enumitem}
\usepackage{xspace}
\usepackage{sansmath}
\usepackage{makecell}
\usepackage[justification=centering]{caption}
\usepackage[ruled,linesnumbered]{algorithm2e}
\usepackage{tikz}
\usepackage{fp}
\usepackage{algpseudocode}
\usepackage{ragged2e}
\usepackage{mathrsfs}
\usepackage{amsthm}
\usepackage{multirow}
\usepackage{colortbl}
\usepackage{wasysym}
\usepackage{fontawesome}
\usepackage{xcolor}
\usepackage{threeparttable}
\usepackage{etoolbox}
\usepackage{soul} % for text highlighting

\newcommand*\blackcircled[1]{%
  \tikz[baseline=(char.base)]{
    \node[shape=circle,fill=black,text=white,inner sep=0.8pt] (char) {#1};}}
\definecolor{customblue}{RGB}{33, 95, 154}

\definecolor{customred}{RGB}{192, 0, 0}

\newcommand{\huanqi}[1]{\textcolor[rgb]{0, 0, 0}{#1}}

\definecolor{Gray}{gray}{0.85}
\definecolor{LightGray}{gray}{0.92}

\newcolumntype{a}{>{\columncolor{Gray}}c}
\newcolumntype{b}{>{\columncolor{LightGray}}c}

\newcommand{\ie}{\emph{i.e.}\xspace}
\newcommand{\eg}{\emph{e.g.}\xspace}

\newcommand{\SystemName}{\texttt{EmbedGenius}\xspace}

\newcommand{\DatasetName}{\texttt{EmbedTask}\xspace}
\makeatletter
\DeclareMathSizes{10}{9.5}{7}{5} % 调整这里的数值以实现接近 \small 的效果
\makeatother

\newlength\maxlentime
\newcommand\timebar[3][red!40]{%
  \FPeval\result{round((#3-1.5)/#2:4)}%
  \rlap{\textcolor{#1}{\hspace*{\dimexpr-\tabcolsep+.5\arrayrulewidth}%
        \rule[-.05\ht\strutbox]{\result\maxlentime}{.95\ht\strutbox}}}%
  \makebox[\dimexpr\maxlentime-0.2\tabcolsep+\arrayrulewidth][r]{#3}}

\def\headertime{New}
\definecolor{headerColor}{RGB}{173, 216, 230}
\definecolor{rowColor1}{RGB}{245, 245, 245}
\definecolor{rowColor2}{RGB}{224, 224, 224}

\author[Huanqi Yang$^{1}$, Mingzhe Li$^{1}$, Mingda Han$^{2}$, Zhenjiang Li$^{1}$, Weitao Xu$^{1}$]{ Huanqi Yang$^{1}$, Mingzhe Li$^{1}$, Mingda Han$^{2}$, Zhenjiang Li$^{1}$, Weitao Xu$^{1}$}
\affiliation{%
\institution{$^1$City University of Hong Kong, $^2$Shandong University}
\country{}
\
}

\begin{document}

\title{EmbedGenius: Towards Automated Software Development for Generic Embedded IoT Systems}
%EmbedGenius: Automating Embedded System Development for General IoT Applications
% \SystemName: Embedded System Development Automation for General IoT Tasks}
% Revolutionizing Embedded System Development with LLM-Powered Automation for General IoT Tasks

\begin{abstract}
% Embedded systems support billions of Internet of Things (IoT) applications essential for interacting with the physical world. 
% However, a fundamental challenge remains in embedded system development, as it is often manually intensive, time-consuming, and prone to errors, especially with complex designs.
% \huanqi{However, a fundamental challenge remains in embedded system development, as the complex processes often necessitate direct developer involvement, making it labor-intensive, time-consuming, and error-prone.}
\huanqi{Embedded IoT system development is crucial for enabling seamless connectivity and functionality across a wide range of applications. However, such a complex process requires cross-domain knowledge of hardware and software and hence often necessitates direct developer involvement, making it labor-intensive, time-consuming, and error-prone.}
To address this challenge, this paper introduces \SystemName, \huanqi{the first fully automated software development platform for general-purpose embedded IoT systems. }
The key idea is to leverage the
reasoning ability of Large Language Models (LLMs) and embedded system expertise to automate the \huanqi{hardware-in-the-loop} development process. 
The main methods include a component-aware library resolution method for addressing hardware dependencies, a library knowledge generation method that injects utility domain knowledge into LLMs, and an auto-programming method that ensures successful deployment. We evaluate \SystemName's performance across 71 modules and four mainstream embedded development platforms with over 350 IoT tasks. Experimental results show that \SystemName can generate codes with an accuracy of 95.7\% and complete tasks with a success rate of 86.5\%, surpassing human-in-the-loop baselines by 15.6\%--37.7\% and 25.5\%--53.4\%, respectively.
We also show \SystemName’s potential through case studies in environmental monitoring and remote control systems development.
\end{abstract}

\maketitle
\settopmatter{printfolios=true}
\pagestyle{plain} % remove running header

\section{Introduction}
\label{sec:intro}
\subsection{Background and Motivation}
% With the proliferation of the Internet of Things (IoT), embedded systems equipped with advanced sensors and actuators have become essential for interacting with the physical world. 
% These systems, utilized for real-world applications, rely on sophisticated development to function effectively. 
% For instance, in a smart agriculture system, sensors measure soil moisture and temperature, while actuators control irrigation systems based on this data~\cite{bwambale2022smart}. Consequently, developing embedded systems is key to enabling IoT applications.
% Embedded systems are the foundation of the Internet of Things (IoT), enabling IoT devices to perform real-world tasks. 
\huanqi{Embedded IoT systems support numerous applications, with the market projected to reach \$258.6 billion by 2032~\cite{precedenceresearch2024}.}  These systems play key roles across various industries, including healthcare~\cite{pham2022pros,truong2020painometry}, agriculture~\cite{dang2022iotree,liller2023towards,vasisht2017farmbeats}, and smart cities~\cite{shi2024soar,santos2018portolivinglab}. For example, in a smart city~\cite{neelakandan2021iot,revathy2017automation}, embedded systems control street lighting and traffic signals based on sensor data to optimize energy use and traffic flow.
% However, their successful application hinges on a highly complex development process, which involves manually resolving dependencies, coding, debugging, and deploying.

\begin{figure}
    \centering
    \includegraphics[width=0.43\textwidth]{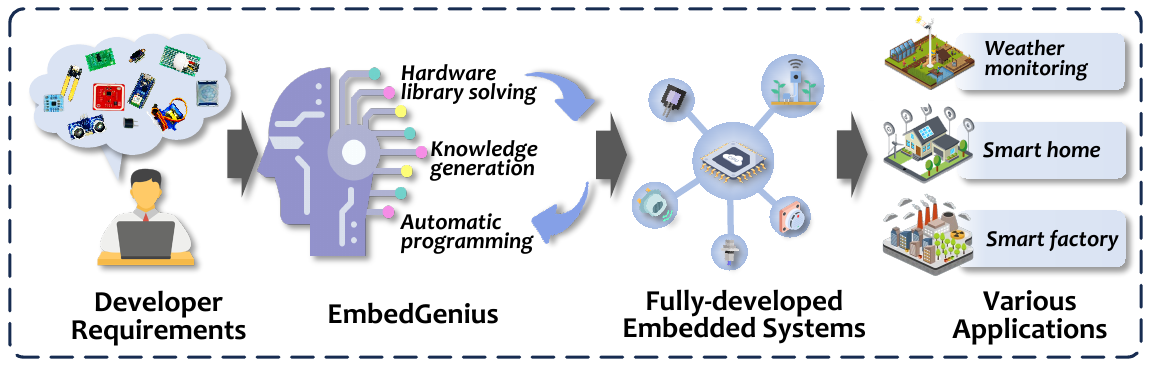}
    \caption{\textbf{Motivation for \SystemName.} \huanqi{A fully automated platform for developing embedded IoT systems (Project: \href{https://embedgenius3.github.io/}{https://embedgenius3.github.io/}).}} 
    % without manual intervention.}}
    \label{fig:intro}
    \vspace{-0.2in}
\end{figure}

\huanqi{Developing such systems straddles the hardware-software interface, demanding a cross-domain understanding of how devices interact with the physical world~\cite{jimenez2013introduction}. Specifically, this intricate process involves several labor-intensive steps.}
% Despite its potential, the process of developing embedded systems is inherently constrained by several labor-intensive steps.
Initially, developers must manually address dependencies by installing and configuring the essential libraries for hardware modules. Next, they write code that defines how the system behaves, considering the functions and attributes of various sensors, actuators, and communication modules.
% Next, they write code to define the device's behavior, considering interactions with various sensors, actuators, and communication modules. 
Afterward, the code is compiled, which often reveals configuration or dependency errors requiring further refinement. Once the code is successfully compiled, it needs to be uploaded to the development platform for testing and deployment. 
Currently, the development of embedded systems predominantly relies on manual processes using traditional Integrated Development Environments (IDEs) (\eg, Eclipse, Keil µVision, and IAR). 
However, the variety and complexity of programming tools lead to high learning and development costs for developers~\cite{programming2006,SaMSolutions2024}. Additionally, the diversity of hardware modules (\eg, sensors, displays, and communication modules) further increases the complexity, making manual development even more challenging. In summary, the traditional embedded IoT system development process is labor-intensive, time-consuming, and error-prone.

This paper introduces a novel automation approach that leverages Large Language Models (LLMs) to streamline embedded IoT system development. Building on the recent success of LLMs (\eg, GPT-4 and Claude-3.5), we aim to harness their capabilities to simplify the development process by automating processes such as dependency-solving, coding, and deployment.
Although there are some existing coding automation tools, such as those utilizing LLMs for general coding tasks~\cite{chen2023codet,zhong2024ldb,ni2023lever}, they only provide partial relief by assisting in tasks like code generation and debugging. For instance, CODET~\cite{chen2023codet} aids in generating code with built-in tests, while LDB~\cite{zhong2024ldb} verifies runtime execution step by step to handle complex logic flow and data operations.
However, they fall short in driving specific hardware devices (\eg, microcontrollers and sensors) due to a lack of hardware-specific knowledge in embedded systems, such as peripheral interface configurations and library dependencies.
% Consequently, these tools cannot fully address the intricacies and specialized requirements of embedded system development. 
Moreover, recent work~\cite{englhardt2024exploring} evaluates how LLMs can assist in embedded system development. However, the study primarily focuses on human-in-the-loop user studies with a limited range of tasks and modules, without reaching automation.
% To address this gap, our approach offers an LLM-powered automation system specifically designed for embedded system development. Fig.~\ref{fig:intro} shows that our system can automate the development process of the embedded system as required by the developer.
To address this gap, we propose an LLM-powered automation system that streamlines the dependency-solving, programming, and deployment processes in embedded system development, resulting in fully-developed embedded systems for various IoT applications as illustrated in Fig.~\ref{fig:intro}.

\subsection{Challenges and Contributions}
We need to tackle several key challenges to achieve the aforementioned goal.

\textbf{Challenge 1: Diversity in Hardware Dependency.} 
Various hardware components, such as communication modules, sensors, and displays, are built on different architectures and perform specialized tasks. Each component relies on specific library dependencies to function effectively, leading to unique challenges in dependency resolution. As noted on the Arduino library website~\cite{arduino_library}, there are over 7,000 libraries across diverse categories, including 1,527 communication libraries, 1,285 device control libraries, 1,435 sensor libraries, and so on. Therefore, accurately identifying and selecting the essential libraries for different hardware components is a significant challenge.
To address this issue, we explore the fundamental principles of library selection, revealing that different libraries exhibit distinct compatibility with specific models and varying support for development board architectures. Additionally, we find that a high version iteration frequency suggests active maintenance and ongoing improvements, reflecting the library's stability and functionality.
% Based on these findings, we evaluate libraries according to their compatible model names and architecture support to ensure the appropriate library selection.
% for effective dependency resolution.
Based on these findings, we propose an automated dependency solving method that can efficiently identify the most suitable libraries for specific hardware components.

\begin{figure}
\centering
\subfigure[File count.]{
\begin{minipage}[t]{0.48\linewidth}
\centering
\includegraphics[width=0.99\linewidth]{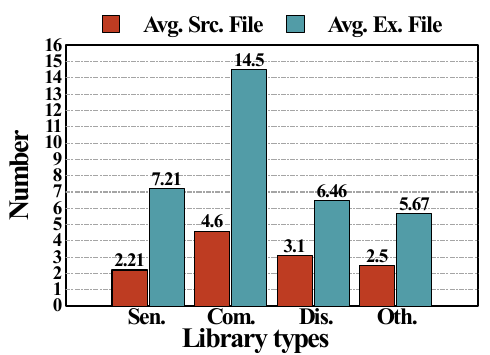}
\label{fig:embed_intro2}
\vspace{-5mm}
\end{minipage}%
}
\hspace{-0.1in}
\subfigure[API count.]{
\begin{minipage}[t]{0.48\linewidth}
\centering
\includegraphics[width=0.99\linewidth]{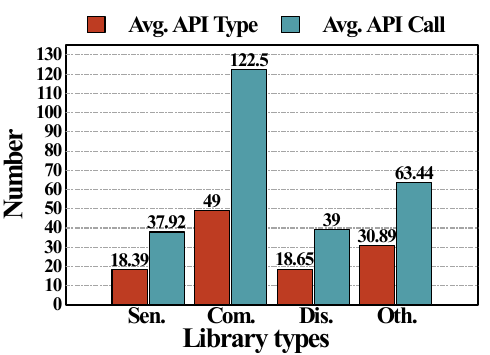}
\label{fig:Embed_intro1}
\vspace{-5mm}
\end{minipage}%
}
\vspace{-2mm}
\centering
\caption{\textbf{Library complexity distribution.} (a) Source and example files, and (b) API types defined in source and API calls in the example. (Sen., Com., Dis., Oth.: Sensors, Communication modules, Displays, Others.)}
\label{fig:lib_com}
\vspace{-0.25in}
\end{figure}

\textbf{Challenge 2: Lack of Library Knowledge.} 
Embedded systems development requires specialized library knowledge, which standard LLMs are not inherently equipped to handle. Moreover, the diversity of libraries across different modules (\eg, Adafruit SSD1306 library for OLED displays or the FastLED library for LED strips~\cite{adafruit_ssd1306_library,fastled_library}) and the variation in their usage further complicate the issue. As shown in Fig.~\ref{fig:lib_com}, different libraries include numerous source and example files that enhance developer understanding by providing extensive API definitions and usage examples. 
This underscores the need for a robust method to enhance LLMs with library knowledge.
Specifically, different libraries often have distinct APIs, each with varying parameters, return values, and usage patterns. Correctly interfacing with these libraries requires a deep understanding of their specific designs. 
% Standard LLMs may generate code that is syntactically correct but contextually inappropriate due to a lack of specialized library knowledge.
To address this challenge, we propose a knowledge generation method that extracts and injects library API and utility knowledge into the LLM’s memory, enabling syntactically and contextually appropriate programming solutions.
% , bridging the gap between general coding practices and the specific demands of embedded IoT systems.

\begin{figure}
    \centering
\includegraphics[width=0.42\textwidth]{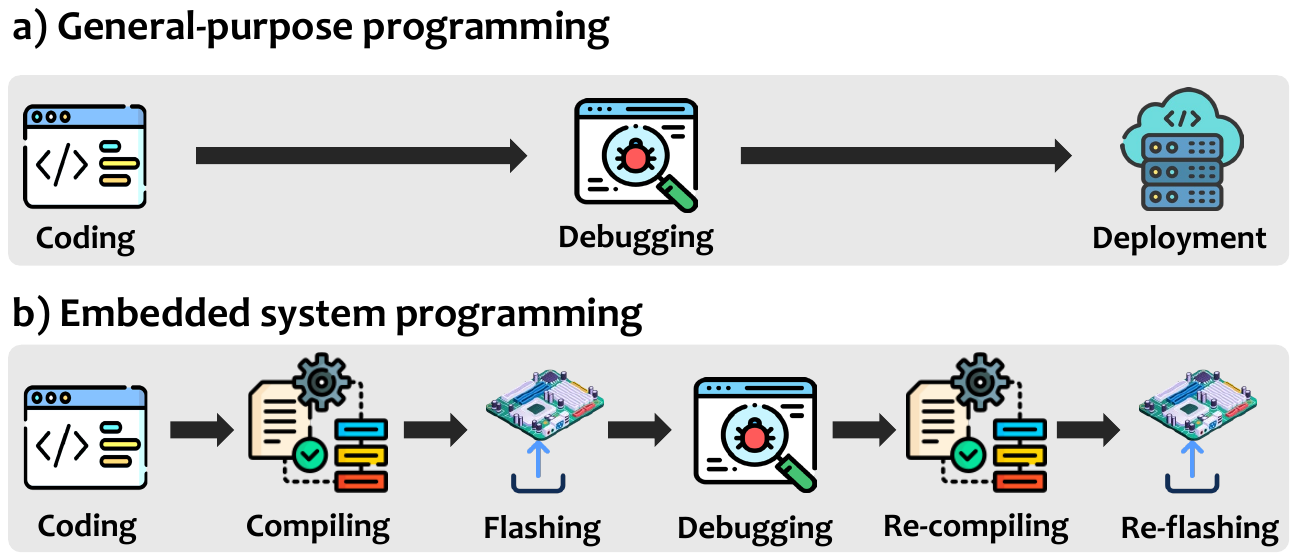}
    \caption{\textbf{Programming pipeline comparison.}}
    \label{fig:pipeline_compare}
    \vspace{-0.3in}
\end{figure}

\textbf{Challenge 3: Complexity of Embedded System Programming.} 
Programming is a crucial part of embedded systems development, involving coding, compiling, and flashing, with each step requiring precise execution~\cite{bayle2013c,evans2011beginning}. Fig.~\ref{fig:pipeline_compare} illustrates the differences between general-purpose programming and embedded system programming. In general-purpose programming, the workflow typically involves coding, debugging, and deployment. In contrast, embedded system programming introduces additional steps such as compiling and flashing, which require specialized configurations and are particularly error-prone.
For example, using an incorrect microcontroller configuration or selecting an incompatible library can result in compilation failures, while coding errors may lead to functional verification failures after flashing.
% Therefore, the inherent complexity of these steps demands an advanced programming mechanism.
To tackle this challenge, we propose an automated programming process that incorporates two nested reasoning and acting feedback loops: a compile loop and a flash loop. Specifically, the compile loop ensures that the correct configurations and libraries are applied before each flashing cycle, while the flash loop verifies that the embedded system functions accurately and completely as described in the user tasks.
% \mingzhe{Fig.xx illustrates the differences between embedded system development and traditional software development, highlighting the additional layers of complexity and hardware-specific considerations involved in the process.}

By incorporating the above solutions, we design and implement \SystemName
% ~\footnote{{Project website:} https://embedgenius3.github.io/ }
, a comprehensive framework that fully automates the dependency-solving, programming, and deployment processes for embedded system development. This advancement lowers the barrier to entry for embedded system development, pushing it into practical, real-world applications. Our extensive evaluation, involving over 70 hardware modules, four development platforms, and over 350 IoT tasks, demonstrates that \SystemName achieves an average coding accuracy of 95.7\% and an average completion rate of 86.5\%, highlighting the effectiveness of \SystemName. 
% We present two case studies to demonstrate how the \SystemName benefits two classical applications: 1) environmental monitoring system, which collect environment data via a sensor and transmit via a LoRa transceiver, and showing the status a OLED display. and 2) remote control system which requires front-end development, including displaying sensor status on a web page via WiFi module and enabling remote device control via relay and PIR sensor.
We present real-life case studies to demonstrate how \SystemName benefits the development of complex systems: 1) an environmental monitoring system and 2) a remote control system. The results show that \SystemName can develop these systems in \SI{2.6}{\minute} and \SI{3.1}{\minute}, respectively.

Our contributions are summarized as follows:
\begin{itemize}[leftmargin=*]
    \item To the best of our knowledge, \SystemName is the first fully automated embedded system development tool that requires no manual intervention, significantly reducing development time and minimizing errors.

    \item We introduce a novel suite of methods, including a library resolution method for hardware dependency solving, a knowledge generation method that enhances LLMs with specialized library knowledge, and an auto-programming method to ensure successful deployment.
    
    % \item We present the \DatasetName dataset, an extensive IoT task dataset with 355 tasks across various applications and difficulty levels, \huanqi{which will be made publicly available upon paper acceptance to support future research.}

   \item We evaluate \SystemName's performance using \DatasetName and various hardware modules. Experimental results show that \SystemName achieves an average coding accuracy of 95.7\% across various IoT tasks. We also conduct real-life case studies to show \SystemName’s potential.

\end{itemize}

\section{Preliminaries}
\subsection{Embedded System Development}
\begin{figure}
    \centering
\includegraphics[width=0.4\textwidth]{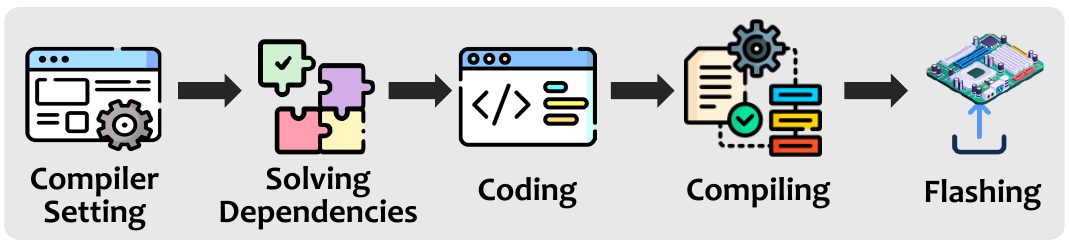}
    \caption{\textbf{Embedded system development pipeline.}}
    \label{fig:pipeline}
    \vspace{-0.3in}
\end{figure}

% The embedded system development process encompasses several integrated steps. Initially, the embedded system is assembled by integrating the development platform $D$ and the necessary modules $M$ (e.g., sensors, communication, and display components). This assembly can be formulated as: $E_s = g(D, M),$ where $E_s$ represents the assembled embedded system. 
Assume an embedded system $E_s = g(D, M)$, where $D$ represents the development platform, and $M$ refers to the connected modules (\eg, sensors, communication modules, and displays). The system $E_s$ is the result of combining $D$ and $M$, forming the foundation for development. The pipeline of embedded system development is shown in Fig.~\ref{fig:pipeline}.

% Once the system is physically assembled, developers configure the hardware settings within the development environment to align with the specific characteristics of the target device. This configuration process can be formulated as: $C_h = f(E_s, P),$ where $C_h$ represents the configured hardware settings, $E_s$ is the assembled embedded system, and $P$ refers to the hardware-level configuration parameters (e.g., peripheral configurations, pin mappings).
\noindent\textbf{Compiler Setting.} The first step in the development process is configuring the settings within the development environment to align with the specific characteristics of the target system $E_s$. This configuration process can be formulated as $C_h = f(E_s, P),$ where $C_h$ represents the configured hardware settings, and $P$ refers to the hardware-level configuration parameters (\eg, pin mappings, target platforms).

\noindent\textbf{Solving Dependencies.} Following the setting, solving dependencies of $E_s$ becomes crucial to ensure smooth integration and functionality. This involves searching for and incorporating the necessary libraries that support the interaction between the development platform and the attached modules.
This process can be expressed as $L_s = S(L_a, E_s),$ where $L_s$ represents the selected libraries that satisfy the dependencies, $L_a$ is the set of available libraries, and the search $S$ is conducted based on the characteristics of $E_s$.

\noindent\textbf{Coding.} Afterward, the developers write the code to implement the desired functionality of the embedded system. Guided by $L_s$ and $C_h$, the coding process is $G = h(T, C_h, L_s),$ where $G$ represents the code, $T$ is the task. 

\noindent\textbf{Compiling and Flashing.} The next step is compiling the code into executable machine code that the hardware can understand. The compilation process is formulated as $B = c(G)$, where $B$ is the resulting binary file. 
Once the code is compiled, the next step is flashing the binary onto the hardware. This step involves transferring the executable machine code onto the embedded system via a suitable interface (\eg, USB, JTAG). The flashing process can be expressed as $F = f(B, E_s),$ where $F$ represents the flashed system.
% After flashing, the debugging phase begins. In this phase, developers test the application on the embedded system, verify that it performs according to the task specifications, and make necessary corrections. The debugging process can be expressed as: $D_d = d(F, T, O),$ where $D_d$ represents the debugging outcome, $O$ is the observed behavior during execution.

Finally, the entire embedded system development process can be expressed as $F = \Phi(E_s, P).$
These processes are typically performed manually, making it labor-intensive, time-consuming, and error-prone

\subsection{Large Language Models}
LLMs typically refer to Transformer-based models~\cite{NIPS2017_3f5ee243} with billions of parameters, trained on extensive text datasets. Examples include models such as ChatGPT~\cite{openai2022chatgpt}, GPT-4~\cite{achiam2023gpt}, PaLM~\cite{chowdhery2023palm}, and LLaMA~\cite{touvron2023llama}, among others. These models demonstrate advanced capabilities not found in smaller models, including mathematical reasoning~\cite{cobbe2021training}, program synthesis~\cite{chen2021evaluating}, and complex multi-step reasoning~\cite{wei2022chain}. An LLM's input is a prompt, which is tokenized into a sequence of tokens, consisting of words or subwords, before processing. The inference process in LLMs can be represented as
\begin{equation}\footnotesize
P(y | x; \theta) = \prod_{t=1}^{T} P(y_t | y_{<t}, x; \theta),
\end{equation}
where \(x\) is the input prompt, \(y = (y_1, y_2, \dots, y_T)\) is the sequence of tokens generated by the model, and \(\theta\) represents the model parameters. The term \(P(y_t | y_{<t}, x; \theta)\) denotes the probability of generating the token \(y_t\) given the previous tokens \(y_{<t}\) and the input \(x\). This probability is computed using a softmax function over the model's output logits:
\begin{equation}\footnotesize
P\left(y_t \mid y_{<t}, x; \theta\right) = \frac{\exp \left(e_{y_t} \cdot h_t\right)}{\sum_{y^{\prime} \in V} \exp \left(e_{y^{\prime}} \cdot h_t\right)},
\end{equation}
where \(e_{y_t}\) is the embedding vector of the token \(y_t\), \(h_t\) is the hidden state at time step \(t\), which is a function of the previous tokens \(y_{<t}\) and the input \(x\), and \(V\) is the vocabulary set.

\begin{figure*}
    \centering
    \includegraphics[width=0.97\textwidth]{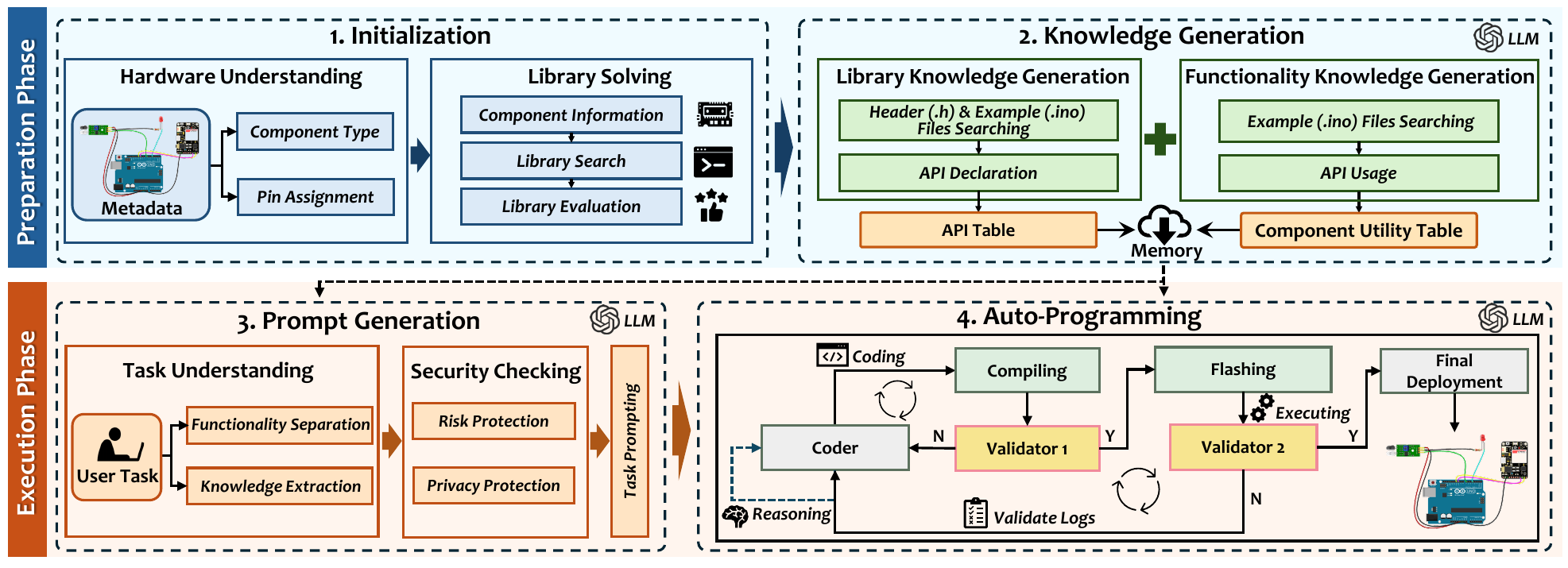}
\caption{\textbf{\SystemName Overview.}}
 % In the preparation phase, \SystemName explores hardware configurations and resolves library dependencies. In the execution phase, it uses memory-augmented LLMs to generate prompts and automate system programming.
    \label{fig:over}
    \vspace{-0.2in}
\end{figure*}

Our system harnesses LLMs' reasoning capabilities to automate embedded system development without human intervention, thereby reducing development time and errors.

\section{System Design}
As shown in Fig.~\ref{fig:over}, \SystemName comprises two phases, preparation and execution phase. During the preparation phase, \SystemName initializes hardware configurations and solves library dependencies, then generates knowledge based on the library content. In the execution phase, \SystemName queries memory-augmented LLMs to generate task prompts and automate system programming. \SystemName employs several techniques to optimize development and improve task completion rates, as detailed below.

% We explain the functioning of \SystemName using the example of automating tasks on an Arduino development board: During the preparation phase, \SystemName explores the board by identifying and interacting with connected components such as LEDs and servos, and records the result in a Hardware Configuration Graph (HCG) memory (Step 1). Next, it traverses all the hardware elements in the HCG and summarizes the tasks they can accomplish (Step 2). During the execution phase, when the user issues a command such as "blink the LED" or "rotate the servo to 90 degrees," the Prompt Generator generates a prompt based on the task, the hardware state description, and relevant information stored in the Hardware Memory. This information includes instructions on how to interact with the specific components. The prompt is then sent to the LLM. Once the LLM provides an answer, the Task Executor parses the action that can be executed on the development board and verifies its security before performing it. If the executor deems the action to be potentially risky, such as causing high current load or short circuits, it will seek confirmation from the user before proceeding. We will explain how \SystemName does all of these in the rest of this section.
\subsection{Initialization}
% Initialization prompting refers to the process of representing underlying hardware metadata information in text and injecting it into the prompt to query the LLM. The goal of initialization prompting is to clearly present the hardware structure information to the LLM and restrict the output of the LLM to generate only valid programs. Additionally, the LLM uses the provided hardware metadata information to complete the compiler settings, ensuring that the code generated is correctly configured for the target hardware. Figure 4 showcases an example of how \SystemName incorporates hardware metadata into a prompt while completing a task.

\subsubsection{Hardware Configuration}
\begin{figure}[h]
    \centering
    \includegraphics[width=0.42\textwidth]{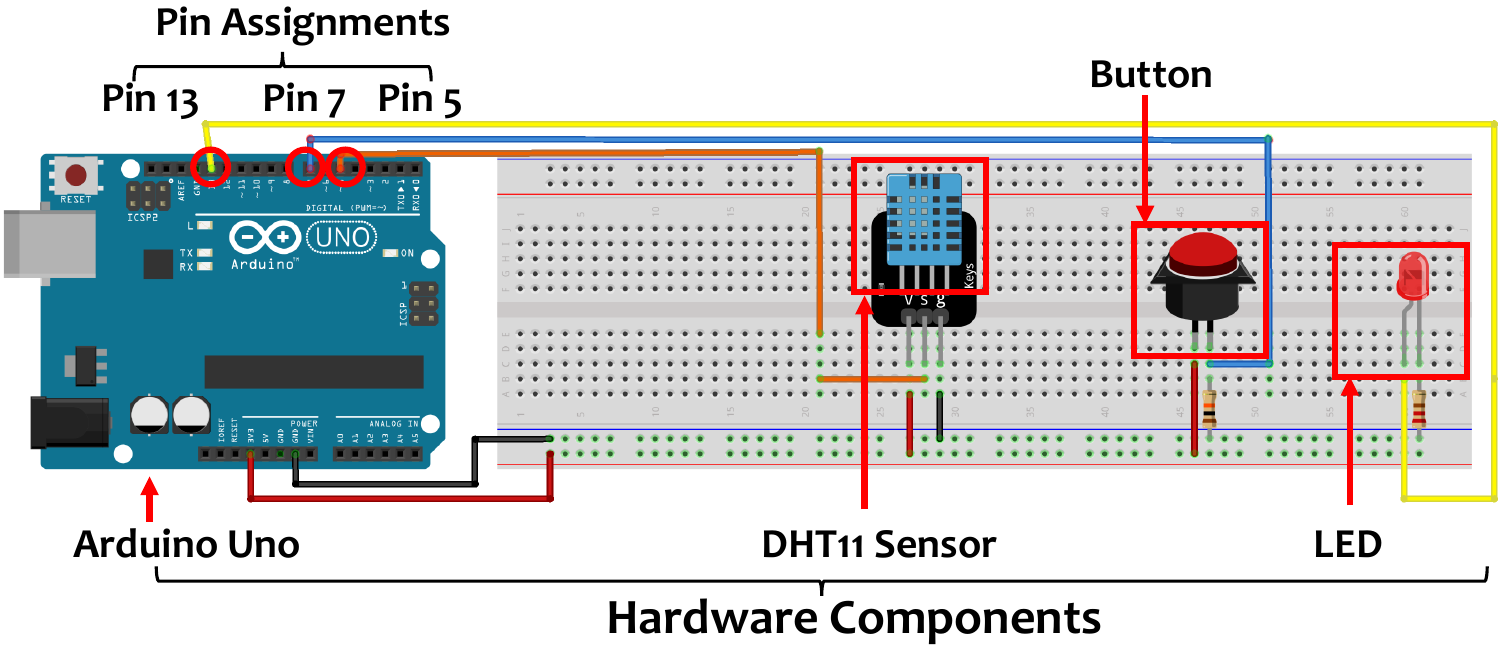}
    \caption{\textbf{Hardware configuration.}}
    \label{fig:hc}
    \vspace{-0.2in}
\end{figure}

% Standardlize input.
% MCU type.
% Component type.
% Component Pinout.
% Explain why these three and we try to simplify as we can. And the final prompt
\SystemName collects and standardizes essential metadata related to hardware information, ensuring seamless integration and minimal setup costs. Users simply input the type of hardware components, including the development platforms $D$ and modules $M$, along with their specific pin assignments $P$. For example, as shown in Fig.~\ref{fig:hc}, a user might specify the development platform as an Arduino Uno, with an LED connected to Pin 13, a button to Pin 7, and a DHT11 sensor to Pin 5. These inputs are the minimal necessary information required from the user, as they depend on the specific choices made for their project. This streamlined interaction minimizes user effort, allowing the system to identify and integrate the components.

\begin{figure*}
    \centering
    \includegraphics[width=0.92\textwidth]{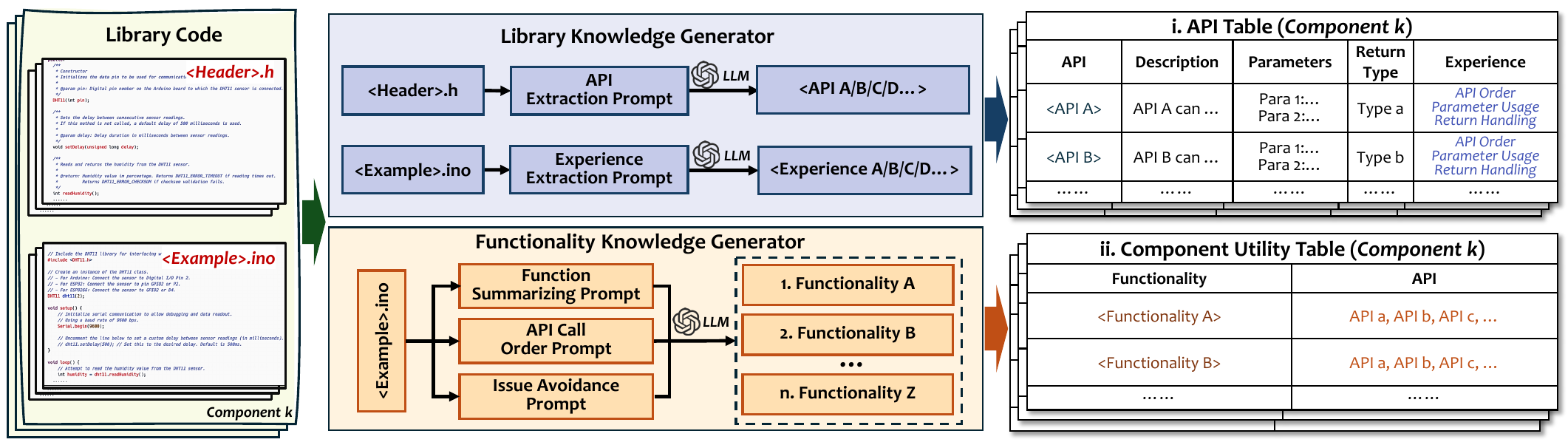}
    \caption{\textbf{Knowledge Generation.}}
    \label{fig:KG}
    \vspace{-0.2in}
\end{figure*}

\subsubsection{Solving Library Dependencies}
\label{subsubsec:solve_lib}
With the provided hardware information, traditional methods rely on users to leverage their experience or manually search for the appropriate library dependencies for each hardware component. However, this approach is inefficient and complex. To address this, we propose an automated library dependency solving method that can efficiently identify the most suitable libraries. The algorithm is shown in Alg.~\ref{alg:lib_selection}. 
\begin{algorithm}[t]
  \scriptsize
  % \SetKwInOut{Input}{Input}\SetKwInOut{Output}{Output}
  \textbf{Initialize:} $comps$, $tgt\_arch$ \textcolor{gray}{// Components list, platform architecture}\\
  % \Output{final\_libs}
  % \BlankLine
  final\_libs $\leftarrow$ \{\}; \textcolor{gray}{// Initialize final libraries}
  
  \For{$c \in$ comps}{
    search\_results $\leftarrow$ \textsc{CLISearch}($c$); \textcolor{gray}{// Perform CLI search}\\
    top\_libs[$c$] $\leftarrow$ \textsc{TopNLibs}(search\_results, $N$); \textcolor{gray}{// Select top N libraries}
  }
  
  \For{$c \in$ comps}{
    best\_lib $\leftarrow \varnothing$; \textcolor{gray}{// Initialize best library}\\
    best\_score $\leftarrow -\infty$; \textcolor{gray}{// Initialize best score}\\
    
    \For{$l \in$ top\_libs[$c$]}{
      details $\leftarrow$ \textsc{ParseLibDetails}($l$); \textcolor{gray}{// Parse library details}\\
      
      % Compute scores
      $m \leftarrow \textsc{CosSim}(details["Name"], c)$; \textcolor{gray}{// Name match score}\\
      $v \leftarrow \frac{\textsc{Count}(details["Vers"])}{\textsc{MaxVerCount}(top\_libs[c])}$; \textcolor{gray}{// Version count score}\\
      $a \leftarrow \textsc{ArchScore}(details["Arch"], tgt\_arch)$; \textcolor{gray}{// Architecture score}\\
      
      % Total score
      $score \leftarrow (m + 0.1 \cdot v + 0.1 \cdot a)\cdot a$; \textcolor{gray}{// Compute total score}\\
      
      \If{$score > best\_score$}{
        best\_score $\leftarrow score$; \textcolor{gray}{// Update best score}\\
        best\_lib $\leftarrow l$; \textcolor{gray}{// Update best library}\\
      }
    }
    final\_libs[$c$] $\leftarrow$ best\_lib; \textcolor{gray}{// Assign best library}
  }
  \Return final\_libs; \textcolor{gray}{// Return final libraries}\\
  \caption{\textbf{Library Selection Algorithm.}}
  \label{alg:lib_selection}
\end{algorithm}
\SystemName begins by extracting component information from the provided hardware metadata. It then performs a search to identify potential libraries, retrieving the top \( N \) libraries for further evaluation.
The evaluation process involves assessing each library based on three criteria:
% \begin{enumerate}[leftmargin=*]
    % \item 
    \textbf{1) Name Match:} This is calculated using cosine similarity between the component name and the library’s name, including its description and any related paragraphs.
    % \item 
    \textbf{2) Version Count:} The score is based on the number of available library versions, normalized between 0 and 1.
    % \item 
    \textbf{3) Architecture Compatibility:} This score is determined by whether the library supports the target hardware architecture. It receives a full score if the library is compatible and zero if it is not.
% \end{enumerate}
Libraries are ranked based on these scores, and the library with the highest overall score is selected for use. If a library's architecture compatibility score is zero, it is deemed unsuitable and excluded from consideration.

\subsection{Knowledge Generation}
\subsubsection{Library Knowledge Extraction}
Once the most suitable libraries are identified, \SystemName needs to learn how to use them. This involves two main steps: API extraction and obtaining API usage information. First, \SystemName identifies the selected library for each component, searches for header (.h) files within the library, and uses the LLM to extract API declarations, summarizing them into the API table. Then, it searches for example (.ino) files within the same library and extracts knowledge on how to use the extracted APIs (e.g., orders, parameters, and return values). This information is further integrated into the API table. As shown in the purple section in Fig. 7, the library knowledge generator extracts the API table content from the header and example files using LLM. 
Given the API extraction prompt \( x_{\text{h}} \) and experience extraction prompt \( x_{\text{ex}} \), the LLM generates API usage tokens \( y_{\text{A}} = (y_1^{\text{A}}, y_2^{\text{A}}, \dots, y_T^{\text{A}}) \) as
\begin{equation}\footnotesize
    P(y_{\text{A}} | x_{\text{h}}, x_{\text{ex}}; \theta) = \prod_{t=1}^{T} P(y_t^{\text{A}} | y_{<t}^{\text{A}}, x_{\text{h}}, x_{\text{ex}}; \theta),
\end{equation}
where \( y_t^{\text{A}} \) is a token in the API sequence, predicted based on previous tokens \( y_{<t}^{\text{A}} \) and the input. This process extracts knowledge about how to effectively utilize the libraries.

\subsubsection{Functionality Knowledge Understanding}
To effectively utilize each component, \textit{\SystemName} creates a component utility table in memory. This table summarizes the functionality and API sequences associated with each component. As shown in the yellow section in Fig.~\ref{fig:KG}, the functionality knowledge generator extracts functionalities from example files by querying the LLM to summarize the API sequences.

Specifically, we summarize the functionality of the examples and identify the API calls related to the component. This forms a utility table, where entries represent the functionalities \( F \) and their associated APIs \( A \). 
Given the functionality understanding prompt \( x_{\text{f}} \), the LLM generates utility tokens \( y_{\text{u}} = (y_1^{\text{u}}, y_2^{\text{u}}, \dots, y_T^{\text{u}}) \) as
\begin{equation}\footnotesize
P(y_{\text{u}} | x_{\text{f}}; \theta) = \prod_{t=1}^{T} P(y_t^{\text{u}} | y_{<t}^{\text{u}}, x_{\text{f}}; \theta).
\end{equation}
After analyzing all example codes, we obtain a utility table in memory containing \( n \) entries, where \( n \) represents the total number of examples. Each entry in the table corresponds to an example and is divided into two parts: \( \langle \text{Functionality}, \text{API} \rangle \).
\textbf{Functionality:} Represents the functionality \( F_i \). It can be perceived as a task that can be completed using these APIs.
\textbf{API:} Represents the sequence of APIs \( A_{i}^{j} \) used from the initial API to the last.
This table provides information about the required operations to achieve each functionality, aiding \SystemName in planning how to complete a given task.

\subsection{Prompt Generation}

\begin{figure}[t]
    \centering
\includegraphics[width=0.41\textwidth]{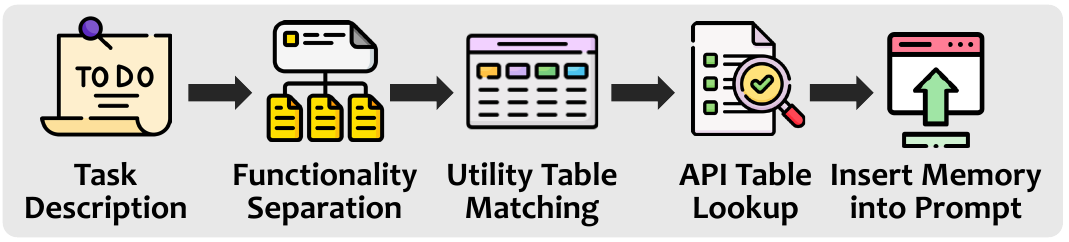}
    \caption{\textbf{Selective memory pick-up process.}}
    \label{fig:memory_inject}
    \vspace{-0.3in}
\end{figure}
\subsubsection{Task Understanding and Memory Pick-up}
\label{subsubsec:selective_memory}
% To leverage domain-specific knowledge from the selected libraries, if we directly incorporate relevant memory into the prompt to guide the LLM. However, this approach can exceed the token limit, such as 4097 tokens for GPT-3.5. Therefore, we selectively include only the most relevant information needed to complete the user's task. Therefor, we design a xxxxxx. The workflow is shown in Fig.xxx, and the detail is as follow.
% To leverage domain-specific knowledge from the selected libraries, if we directly incorporate relevant header or example file content into the prompt to guide the LLM, this approach can exceed the token limit, such as 4097 tokens for GPT-3.5. Also, this will incur more cost. Therefore, we selectively include only the most relevant information needed to complete the user's task. Therefore, we design a memory pick-up method for xxxx. The workflow is shown in Fig.1, and the detail is as follows.
To leverage memorized knowledge from chosen libraries, directly embedding pertinent header or example file content into the prompt to guide the LLM may surpass the token limit, such as the 4096 tokens for GPT-4. Moreover, this approach can escalate costs. Hence, we opt to incorporate only the most crucial information necessary for accomplishing the user's task. To address this, we devise a selective memory pick-up method. The workflow is shown in Fig.~\ref{fig:memory_inject}, with details below.
\noindent\textbf{Functionality Separation.} 
We query the LLM to separate the task description into functionalities corresponding to each component. For instance, given the task "Record the DHT11 temperature reading to SD card," the identified functionalities would be: "initialize DHT11 sensor," "initialize SD card," "read temperature from DHT11 sensor," and "store data to SD card."
\noindent\textbf{Knowledge Extraction.} 1) Utility Table Matching:
For each functionality in the task, we match it to functionalities in the memory's utility table by comparing their similarity.
% We use a TF-IDF (Term Frequency-Inverse Document Frequency) method~\cite{aizawa2003information} that represents natural language sentences as weighted vectors, where sentences with similar meanings have more closely aligned vector representations based on the significance of their terms. The cosine similarity between the vectors of the function in memory \( St \) and the current function \( Ct \) is denoted as \(\operatorname{sim}(E(St), E(Ct))\). We can then find the \( k \) most similar tasks in the memory, denoted as \(\{St_1, St_2, \ldots, St_k\}\). For each \( St_i \), we retrieve the corresponding function, library, and functional API from the component function table.
We use the Term Frequency-Inverse Document Frequency (TF-IDF) method~\cite{aizawa2003information} to represent natural language sentences as weighted vectors. The cosine similarity between the vectors of the functionality in utility table \( F_t \) and the current functionality \( C_t \) is denoted as \(\operatorname{sim}(E(F_t), E(C_t))\). We then identify the \( k \) most similar functionalities in the memory, denoted as \(\{F_{t1}, F_{t2}, \ldots, F_{tk}\}\). For each \( F_{ti} \), we retrieve the corresponding APIs from the utility table.
2) API Table Lookup:
Next, we look up the API table to retrieve usage information about the matched APIs. This information \(A_T\) is then incorporated into the prompt, allowing the LLM to leverage this context for more accurate and relevant code generation.

% \noindent\textbf{Prompt Generation:} use lookuped information and user task and Hardware Configuration information to structure the fininal prompt for coding.

\subsubsection{Security Checking.}
% \textbf{1) Risk Protection.} Certain actions in embedded systems may potentially alter local or server data or be irreversible once performed. These actions are considered risky and require user confirmation before execution. For example, updating the firmware of a smart thermostat can lead to significant changes and cannot be undone once performed.
% To mitigate risks, the system prompts the user for confirmation before executing risky actions. For instance, before sending the command to change the Wi-Fi credentials of a networked camera, the system asks, \textit{“This action will change the Wi-Fi settings of your camera. If this action potentially leads to a change of device state or server data that requires user confirmation, please answer requires\_confirmation=Yes.”} Additionally, key phrases such as “warning” are utilized in the CLI to identify further and flag potentially risky actions. After receiving confirmation from the user, the system proceeds with the action.
\textbf{1) Risk Protection.} During embedded system development, certain actions may lead to system malfunctions or disrupt normal device operation. These actions are considered risky and require developer confirmation. For example, a developer may adjust the PWM signal frequency to control a motor's speed. If the frequency is set incorrectly, it could cause the motor to overheat or even get damaged.
To mitigate such risks, the system prompts the developer for confirmation before executing these high-risk operations. The system will ask: \textit{"This action could lead to overheating or potential damage to the motor, please answer requires\_confirmation=Yes."} Additionally, key phrases like "warning" are jumped to flag potentially risky actions. The system proceeds with the action only after receiving confirmation from the developer.
\textbf{2) Privacy Protection.} It is crucial to protect sensitive information such as device IDs, network credentials, and user names. We add a privacy filter that can mask private information in user queries.
For example, if a user query includes \textit{“Set device ID to 12345 and Wi-Fi password to ‘password123’,”} the system detects this Personal Identifiable Information (PII). The PII is then replaced with non-private placeholders before sending the query to the cloud, resulting in \textit{“Set device ID to <device\_id> and Wi-Fi password to <password>.”} After receiving the response from the cloud, the system maps the placeholders back to the original sensitive information and executes the action. 
% This ensures that sensitive information is protected.

\subsubsection{Task Prompting.}
\label{subsubsec:prompt}
% We combine the extracted API information, user task description, and hardware metadata to create the final prompt. The prompt includes:
% \begin{itemize}[leftmargin=*]
%     \item \textbf{Task(\(T\)):} User-defined task.
%     \item \textbf{Hardware metadata (\(H\)):} Configurations for the involved hardware components.
%     \item \textbf{Extracted Knowledge (\(A\)):} Relevant API usage information from memory.
%     \item \textbf{Contextual Info (\(C\)):} Additional information like goals, constraints, and coding rules.
% \end{itemize}
% This structured prompt ensures the LLM generates accurate and optimized code while staying within token limits and reducing costs.
% To effectively guide the LLM in generating code, we construct a structured prompt that includes the user-defined task \(T\), configuration metadata for the involved hardware components \(H\), relevant API usage information \(A_T\), and coding rules \(C\). This structured prompt ensures the LLM generates accurate and optimized code while staying within token limits and reducing costs.
To effectively guide the LLM in generating code, we construct a structured prompt that includes the user-defined task, configuration metadata for the involved hardware components, relevant API usage information, and coding rules. This structured prompt ensures the LLM generates accurate and optimized code while staying within token limits and reducing costs.

% \subsubsection{Compiler Setting}
% \SystemName uses Arduino CLI, command line, and Python scripts for automatically setting the compiler and executing the necessary tasks to compile and upload the code to the development platform. The Arduino CLI provides functions to configure the board settings, manage libraries, and compile and upload sketches. Specifically, Arduino CLI can be used to set the target board, specify the port, install and update required libraries, and compile the code.
% The typical workflow involves several steps: \textbf{1) Setting the Target Platform:} Use commands like `board attach -b arduino:avr:uno` to specify. \textbf{2) Installing Libraries:} Commands such as `lib install "DS18B20"` install necessary libraries. \textbf{3) Compiling the Code:} The `compile --fqbn arduino:avr:uno path/to/sketch` command compiles the sketch for the specified board. \textbf{4) Uploading the Code:} Finally, the `upload -p COM3 --fqbn arduino:avr:uno path/to/sketch` command uploads the compiled code to the hardware.
% By integrating the above tool, \SystemName automates the development for embedded systems, streamlining the workflow and reducing the need for manual intervention.

\subsection{Auto-Programming}
The auto-programming method plays a crucial role in our system, guaranteeing that the produced code is syntactically correct and functionally accurate. It functions within two nested dynamic reasoning and action loops, as shown in Fig.~\ref{fig:over}. 
The algorithm design is detailed in Alg.~\ref{al:autodebug}.

\begin{algorithm}[tbp!]
  \scriptsize

  \textbf{Initialize:}
  $T, M_c, M_e, {M_{v1}},{M_{v2}}$; \textcolor{gray}{// Task, Coder, Executor, two Validators}\\
  
  Set $FL_c, FL_f \leftarrow []$; \textcolor{gray}{// Initialize feedback Log}\\
  Set $t_f = 0$; \textcolor{gray}{// Initialize flash debug trial count}\\
\Repeat{$FL_f == Success$ or $t_f \geq \text{max compile debug trials}$}{
Set $t_c = 0$; \textcolor{gray}{// Initialize compile debug trial count}\\
  \Repeat{$FL_c == Success$ or $t_c \geq \text{max flash debug trials}$}{
    \textbf{Generate Code:} \\
    $G \leftarrow M_c(T, FL_c, FL_f)$; \textcolor{gray}{// Generate code with DEBUG INFO inserted}\\% Generate code and DEBUG INFO with feedback
    \textbf{Compile Code:} \\
    $O_c, B \leftarrow M_e(G)$; \textcolor{gray}{// Compile code, get output and binary file}\\
    $t_c \leftarrow t_c + 1$; \textcolor{gray}{// Increment trial count}\\
    \textbf{Validate Compile:}\\
    $FL_c \leftarrow {M_{v1}}(O_c)$;\\
    % \If{$FL_c \neq []$}{
    % \textbf{continue}; \textcolor{gray}{// Skip to the next iteration of compile loop}
    % }
    }
    \If{$FL_c == Success$}{
    \textbf{Flash Code:} \\
    $O_f \leftarrow M_e(B)$; \textcolor{gray}{// Flash the binary file and get the debug output}\\
    
    \textbf{Validate Execution:} \\
      $FL_f \leftarrow M_{v2}(O_f,T)$  \\ %$[FL_f,e_{f} \leaftarrow M_{vf}(L_t,T)$
      % \If{$FL_f \neq []$}{
      %   \textbf{continue}; \textcolor{gray}{// Skip to the next iteration of flash loop}
      % }
    % }
    $t_f \leftarrow t_f + 1$; \textcolor{gray}{// Increment trial count}\\
    }
%     \Else{
%     \Return{[]}  \\
% }
    
    }
    $G_f \leftarrow G - D(G)$; \textcolor{gray}{// Clean final code}\\
    \Return{$G_f$}
  \caption{\textbf{Auto-Programming Algorithm.}}
  \label{al:autodebug}
\end{algorithm}

\begin{figure*}[tbp!]
    \centering
    \includegraphics[width=0.98\textwidth]{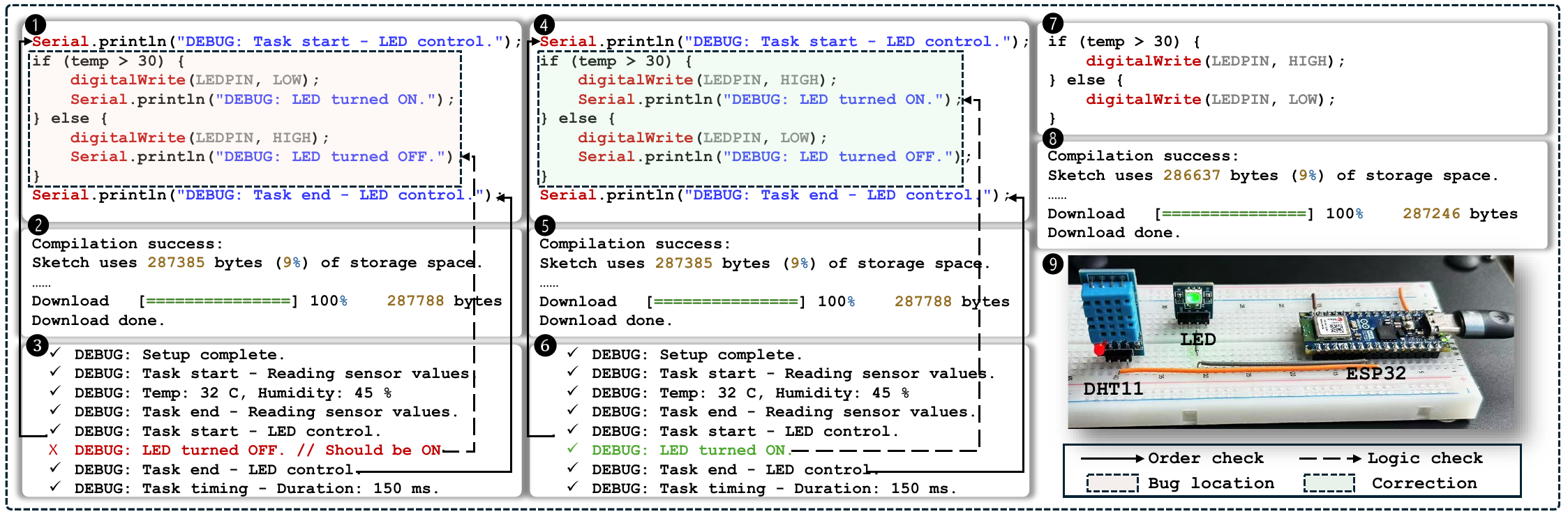}
    \caption{\textbf{Auto-programming example.}}
    \label{fig:de}
    \vspace{-0.2in}
\end{figure*}

% \subsubsection{Coder.} The process initiates with the Coder, which receives tasks from the Prompt Generator and accesses the knowledge memory to generate code. Guided by our prompt as introduced in Sec.~\ref{subsubsec:promtp}, the Coder integrates DEBUG INFO into the code to monitor various aspects of the execution state, including the start and end of each task, their timing, and sequence. This integration is crucial as it allows for the immediate detection and correction of errors, ensuring that similar mistakes do not recur.
% Let $G$ represent the generated code, and $D(G)$ be the set of DEBUG INFO annotations. 
% For each task $T_i$ in the generated code $G$, DEBUG INFO statements $D_i$ are included such that $D(G) = \bigcup_{i=1}^{n} D_i$, where $n$ is the total number of tasks. These annotations provide information about the start time $t_{start}$, end time $t_{end}$, and execution order $o_i$ of each task.
% \subsubsection{Coder.} The Coder begins by receiving a task prompt from the Prompt Generation process to generate code. 
\subsubsection{Coder.} The Coder begins by receiving a task prompt from the Prompt Generation process to generate code. This process can be represented by 
\begin{equation}\footnotesize
P(G | x_T; \theta) = \prod_{t=1}^{T} P(G_t | G_{<t}, x; \theta),
\end{equation}
where \(G\) represents the generated code, \(x_T\) is the task prompt, and \(\theta\) denotes the model parameters. 
% The term \(P(G_t | G_{<t}, x; \theta)\) denotes the probability of generating the token \(G_t\) in the code given the previous tokens \(G_{<t}\) and the task prompt \(x_T\). 
Guided by the prompt described in Sec.~\ref{subsubsec:prompt}, the coding rules include integrating DEBUG INFO into the code to monitor various aspects of the execution state, such as order and logic. To facilitate automated monitoring, DEBUG INFO is dynamically generated by the LLM in conjunction with code generation. 
Let \(D(G)\) represent the set of DEBUG INFO annotations. For each subtask \(T_i\) in \(G\), DEBUG INFO statements \(D_i\) are included such that \(D(G) = \bigcup_{i=1}^{n} D_i\), where \(n\) is the total number of subtasks. 
% These annotations provide information about the start time \(t_{start}\), end time \(t_{end}\), and execution order \(o_i\) of each subtask.

% For each task $T_i$ in the generated code $G$, DEBUG INFO statements $D_i$ are included such that:
% \begin{equation} \small
%     D(G) = \bigcup_{i=1}^{n} D_i,
% \end{equation}
% where $n$ is the total number of tasks. These annotations provide information about the start time $t_{start}$, end time $t_{end}$, and execution order $o_i$ of each task.

\subsubsection{Compile Loop.} The code is then passed to the compiler, which checks for syntax errors, optimizes the code, and converts it into machine-readable instructions. Afterward, the compilation result is sent to query the Compile Validator, which queries the LLM to check if the compilation was successful. If unsuccessful, it returns summarized logs to the Coder for error correction. If successful, the code is deemed ready for flashing.

\subsubsection{Flash Loop.}  
% Then the compiled file is flashed to the development platform, where the code can be executed. Afterwards, Flash Validator uses the embedded DEBUG INFO to identify and correct errors by performing two checks:
% \begin{enumerate}[leftmargin=*]
%     \item \textbf{Order Check:} Ensuring that for each subtask $T_i$, the start time is before the end time ($\forall T_i, \; t_{start}(T_i) < t_{end}(T_i)$), and that the execution order follows the expected sequence.
%     % ($\forall T_i, \; o(T_i) \text{ follows the expected order}$).

%     \item \textbf{Logic Check:} Verifying that the internal logic of each subtask $T_i$ adheres to the expected behavior and constraints, ensuring that the code performs correctly beyond just the order and timing of tasks.
% \end{enumerate}
The compiled file is flashed to the development platform, where the code is executed. The Flash Validator leverages the LLM to utilize the embedded DEBUG INFO to identify and correct errors by performing two checks: the \textbf{Order Check}, which ensures that for each subtask $T_i$, the start time precedes the end time ($\forall T_i, ; t_{\text{start}}(T_i) < t_{\text{end}}(T_i)$) and that the execution order follows the expected sequence; and the \textbf{Logic Check}, which verifies that the internal logic of each subtask $T_i$ adheres to the expected behavior and constraints, ensuring correct functionality beyond just task order and timing.
If discrepancies are detected, the Flash Validator provides feedback to the Coder for code refinement and restarts the loop until the code passes validation.

% Fig.~\ref{fig:hc} shows an example involves reading temperature and humidity data from a DHT11 sensor and controlling an LED based on temperature readings. Initially, the code contains a logical error where the LED is incorrectly turned off when the temperature exceeds 30°C. Embedded DEBUG INFO statements provide detailed logs, including task start times \( t_{start} \), end times \( t_{end} \), execution order \( o_i \), and task duration. For each task \( T_i \) in the generated code \( G \), DEBUG INFO statements \( D_i \) are included such that \( D(G) = \bigcup_{i=1}^{n} D_i \), where \( n \) is the total number of tasks. The Validator uses these logs to detect the discrepancy in the LED control logic and provides feedback to the LLM programmer. The corrected code ensures the LED turns on when the temperature exceeds 30°C and off otherwise, with all tasks executing in the correct order and within expected timings. Finally, the DEBUG INFO is removed, and the error-free code is deployed to the embedded development board.
Fig.~\ref{fig:de} presents an example involving the reading of temperature and humidity data from a DHT11 sensor and the subsequent control of an LED based on the temperature readings. The process is detailed as follows.
The Coder generates the initial code $G$ with embedded DEBUG INFO statements $D(G)$ (step \blackcircled{1}). 
The code is then compiled, and its readiness for deployment is evaluated. If the code is not ready, the Compile Validator identifies the issues and prompts corrections. In this scenario, the code passes the validation successfully (step \blackcircled{2}).
Then it flashs the code to the embedded development board and collects the execution logs. Subsequently, the Flash Validator analyzes these logs to assess the correctness of the code. In this scenario, the Flash Validator identifies a logical error where the LED is incorrectly turned off when the temperature exceeds 30°C. It provides feedback to the Coder for necessary code correction (step \blackcircled{3}).
Upon receiving the feedback, the Coder evaluates and implements the required changes to correct the LED control logic in the code $G$. 
Additionally, the Coder updates the corresponding DEBUG INFO statements $D(G)$ to reflect the modifications. (step \blackcircled{4})
The process is repeated iteratively until the code is verified to be successfully compiled and error-free (step \blackcircled{5}--\blackcircled{6}).
Finally, the DEBUG INFO are removed, and the clean code is redeployed to the development board (step \blackcircled{7}--\blackcircled{9}).
Auto-Programming integrates the LLM programmer, Executor, and Validator to create a robust feedback loop, ensuring the generated code is syntactically and functionally correct.
% , facilitating efficient and reliable hardware development.

\begin{figure*}[tbp!]
    \centering
\includegraphics[width=0.98\textwidth]{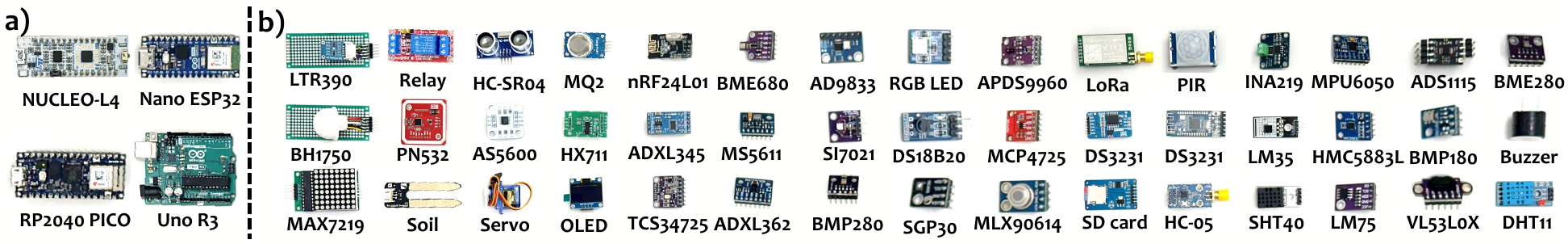}
    \caption{\textbf{Experimental devices.} a) Development platforms and b) hardware modules. Detailed list available at \href{https://embedgenius3.github.io/\#Hardware}{https://embedgenius3.github.io/\#Hardware.}}
    \label{fig:device}
    \vspace{-0.2in}
\end{figure*}

\section{Evaluation}
\subsection{Experimental Setup}

% \textbf{\SystemName implementation.} We employed HTTP requests to interface with LLM APIs, enabling interaction with LLMs such as GPT-4, with GPT4o set as the default model. The automation framework was built using Python 3.9 and integrated with the Arduino CLI compiler.
\textbf{\SystemName Implementation.} The automation framework leverages HTTP requests to interface with LLM APIs, enabling seamless interaction with models such as GPT-4, with GPT-4o set as the default model. It was developed using Python 3.9 and is fully integrated with the Arduino CLI compiler, a tool that offers a universal way to interact with the development environment from the command line.

\noindent\textbf{Development Platforms.}
To comprehensively evaluate the performance of \SystemName, we selected four popular development platforms that are highly representative in consumer electronics and education: Uno R3, NUCLEO-L4, Nano RP2040, and Nano ESP32, as shown in Fig.~\ref{fig:device} (a). The specifications of these platforms are shown in Tab.~\ref{tab:platform}. These boards showcase diverse MCU architectures, varying computational capabilities, and distinct application scenarios, effectively representing the majority of embedded platforms.

\begin{table}[t]
\caption{\textbf{Development platform specifications.}}
\scriptsize
\begin{threeparttable}
\setlength{\tabcolsep}{4pt}
\begin{tabular}{lccccc}
\toprule
\textbf{Platform}&
\textbf{MCU\tnote{*}}&
\textbf{Max. freq.}&
\textbf{FLASH}&
\textbf{SRAM}&
\textbf{Size}
\\ \midrule
Uno R3 & ATmega328P & 20MHz & 32KB & 2KB &68.6$\times$53.4mm\\
NUCLEO-L4& STM32L432KC & 80MHz & 256KB &64KB&50.3$\times$18.5mm\\
Nano RP2040& RP2040 & 133MHz& 16MB &264KB &45.0$\times$18.0mm\\
Nano ESP32& ESP32-S3 & 240MHz & 32MB & 512KB& 43.1$\times$17.8mm\\
\bottomrule
\end{tabular}
\begin{tablenotes}\scriptsize
    \item[*]AT: Atmel, STM: STMicroelectronics, RP: Paspberry PI pico, ESP: Espressif.
\end{tablenotes}
\end{threeparttable}
\label{tab:platform}
\vspace{-0.2in}
\end{table}

\noindent\textbf{Modules.} To demonstrate the extensive compatibility of \SystemName, 71 modules were utilized to validate its functionality across a wide range of embedded devices and modules. As shown in Fig.~\ref{fig:device} (b), we display most of the modules used. Some similar hardware was omitted due to space constraints. These modules encompass environmental sensors (\eg, temperature, humidity, atmospheric pressure), motion sensors (\eg, magnetic angle sensors, accelerometers), communication modules (\eg, Bluetooth, LoRa), display modules, control devices, and others, thereby representing a comprehensive array of devices.
For each module, various tasks of differing complexity levels were employed to assess their integration with \SystemName, as will be detailed below. 
% These tasks will be described in detail in the following section.

\begin{figure}
\centering
\subfigure[Number.]{
\begin{minipage}[t]{0.49\linewidth}
\centering
\includegraphics[width=0.99\linewidth]{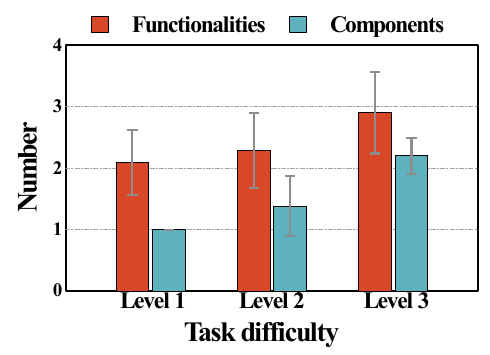}
\label{fig:embed_complex1}
\vspace{-4mm}
\end{minipage}%
}
\hspace{-0.1in}
\subfigure[Complexity.]{
\begin{minipage}[t]{0.49\linewidth}
\centering
\includegraphics[width=0.99\linewidth]{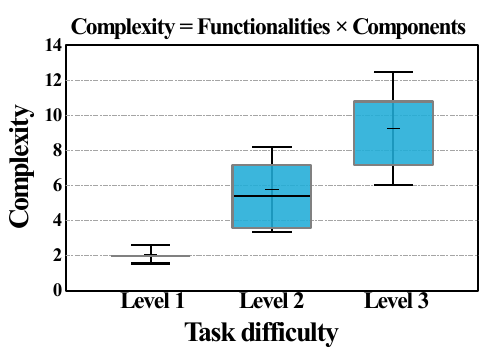}
\label{fig:embed_complex2}
\vspace{-4mm}
\end{minipage}%
}
\vspace{-2mm}
\centering
\caption{\textbf{Complexity of \DatasetName.}}
\label{fig:embed_complex}
\vspace{-0.2in}
\end{figure}

%interface
%同时驱动并协调不同模块以进行多种任务的能力是\systemname通用性的另一体现。->为了说明这一点,我们使用了xxx,xxx以及xxx等，涵盖yy种功能, 总计多达80多种模块来对\system 进行测试。->对于每种module，我们将不同难度等级的tasks来评估\system在该种情况下的可用性。
\noindent\textbf{Task Dataset.}
We introduce \DatasetName, an IoT task dataset designed to evaluate the effectiveness of a fully automated embedded system development process from start to finish. 
% This dataset will be made available, 
% on GitHub~\footnote{Github website} as a JSON file, 
% enabling researchers to reproduce the evaluation environment
% ~\footnote{More details (\eg, demo, code, updates) will be released with this paper.}.
\DatasetName includes 355 tasks, covering different modules and varying complexity levels. These tasks span a range of applications, such as environmental monitoring, motion detection, data communication, digital display, and motion control.
% \DatasetName categorizes tasks into three difficulty levels: simple, moderate, and challenging. Simple tasks evaluate a module's basic functions. Moderate tasks require operating additional simple modules or engaging one more logic. Challenging tasks are more intricate, involving collaboration between multiple modules and engaging more logic. The complexity is defined by number of functionalities multiply number of components engaged.
\DatasetName classifies tasks into three difficulty levels: 
\begin{itemize}[leftmargin=*]
\item \textbf{Level 1 tasks} assess the basic functionality of a single module.  
\textit{Example}: Reading temperature every second and displaying the data on a serial monitor.  
\textit{Components}: DHT11 sensor.  
\textit{Functionalities}: (1) Read temperature, (2) Display data on the serial monitor.

\item \textbf{Level 2 tasks} involve the operation of additional simple modules or the integration of an additional layer of logic.  
\textit{Example}: Reading humidity every second and turning on an LED if humidity exceeds 20\%.  
\textit{Components}: DHT11 sensor, LED.  
\textit{Functionalities}: (1) Read humidity, (2) Compare it with a threshold, (3) Turn on the LED.

\item \textbf{Level 3 tasks} are more complex, requiring collaboration between multiple modules and engaging advanced logic.  
\textit{Example}: Recording temperature data to an SD card every 2 seconds while controlling a servo to rotate 20 degrees counterclockwise each second.  
\textit{Components}: DHT11 sensor, SD card module, Servo motor.  
\textit{Functionalities}: (1) Read temperature, (2) Record data to the SD card, (3) Control the servo motor's rotation.

\end{itemize}
Task complexity is formally defined as the product of the number of functionalities (\(N_f\)) and the number of components (\(N_c\)) involved:  $\text{Task Complexity} = N_f \times N_c$.
% \begin{equation}\small
% \text{Task Complexity} = N_f \times N_c
% \end{equation}
This metric aligns with established system engineering principles, where complexity increases with both functionalities and components~\cite{boehm1984software}.
% This equation ensures that tasks with greater functionalities and a higher number of engaged components are systematically categorized into more complex levels, providing a consistent metric for classification.
% To accommodate the limited interfaces and computational capabilities of various development platforms, challenging tasks are restricted to a maximum of three modules. 
As illustrated in Fig.\ref{fig:embed_complex}, the variation in task complexity arises from differences in the number of functionalities and components across difficulty levels.

\huanqi{Compared with existing benchmark~\cite{englhardt2024exploring}, our experimental setup involves $30\times$ more tasks, $5\times$ more modules, and $2\times$ more development platforms. This coverage ensures that our system is capable of addressing a wide range of IoT tasks with diverse hardware modules.}

\noindent\textbf{Metrics.}
To evaluate \SystemName's performance, we consider a sequence of functionalities $\{F_1, F_2, \ldots, F_m\}$ alongside a sequence of API usages $\{A_1, A_2, \ldots, A_n\}$ performed by human annotators to complete a task $T$. If \SystemName can use a sequence of APIs $\hat{A} = \{\hat{A}_1, \hat{A}_2, \ldots, \hat{A}_n\}
$ corresponding to each functionality, we employ the following two metrics:
\begin{enumerate}[leftmargin=*]
    \item \textbf{Coding Accuracy:} This metric is defined as the ratio of API usage $\hat{A}_i$ that matches the reference $A_i$ defined in library, expressed as $P(\hat{A}_i = A_i)$. An API usage is deemed correct if both the API and its parameters are accurate. 
    
    \item \textbf{Completion Rate:} This is the probability that the system completes all functionalities in task correctly within one attempt, represented as $P(\hat{A} = A)$. This metric indicates the likelihood of successfully completing a task.
\end{enumerate}

% \begin{figure}[t]
%     \centering
% \includegraphics[width=0.31\textwidth]{fig/distribution_broken_y_axis_v10.pdf}
%     \caption{\textbf{The distribution of tasks in \DatasetName across different numbers of functionalities.}}
%     \label{fig:distribution_task}
%     \vspace{-0.3in}
% \end{figure}

\subsection{Overall Performance}
\begin{figure}
\centering
\subfigure[By module type.]{
\begin{minipage}[t]{0.49\linewidth}
\centering
\includegraphics[width=0.99\linewidth]{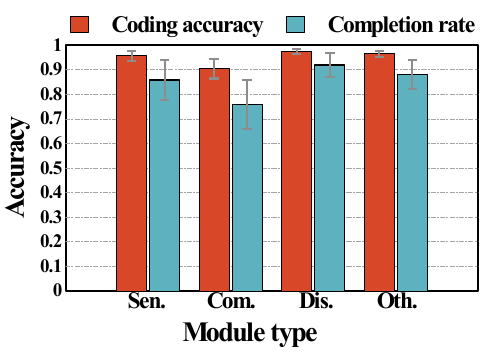}
\label{fig:overall_result1}
\vspace{-4mm}
\end{minipage}%
}
\hspace{-0.1in}
\subfigure[By task difficulty.]{
\begin{minipage}[t]{0.49\linewidth}
\centering
\includegraphics[width=0.99\linewidth]{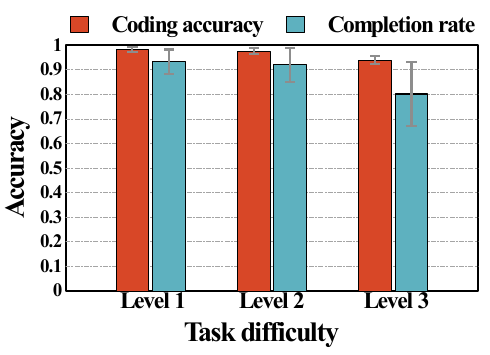}
\label{fig:overall_result2}
\vspace{-4mm}
\end{minipage}%
}
\vspace{-2mm}
\centering
\caption{\textbf{Overall performance.}}
\label{fig:overall_result}
\vspace{-0.2in}
\end{figure}

\noindent\textbf{Overall Accuracy.} 
Fig.~\ref{fig:overall_result} presents the accuracy percentages for task execution across various modules and task difficulty levels using \SystemName. Specifically, coding accuracy is 95.8\%, 90.5\%, 97.6\%, and 96.5\%, with corresponding completion rates of 85.9\%, 76.1\%, 92.2\%, and 88.2\% for the sensor, communication, display, and other module types, respectively. The communication modules demonstrate lower accuracy due to the increased complexity of their associated libraries, as shown in Fig.~\ref{fig:lib_com}.
Across difficulty levels, coding accuracy stands at 98.4\%, 97.6\%, and 93.9\%, with completion rates of 93.3\%, 92.1\%, and 80.2\% from simple to challenging. These results highlight the effectiveness of \SystemName across module types and task complexities.

\noindent\textbf{Comparison with Baselines.} While there is no fully automated method for embedded system development, we compare our approach against three human-in-the-loop methods. The primary baseline is an existing LLM-based prompt design for embedded system development~\cite{englhardt2024exploring}, referred to as LLM-Prompt. This method requires a human to manually extract information from the compiler and provide it as input to the LLM, enabling debugging assistance. Additionally, we use Duinocode, an LLM-based code generator for Arduino, as our second baseline~\cite{duinocode}. The third baseline, referred to as LLM-direct, involves directly prompting the LLM to assist with programming tasks. 
As shown in Tab~\ref{tab:baseline}, \SystemName achieves significantly higher performance, with coding accuracy improvements of 32.4\%, 15.6\%, and 37.7\% and completion rate increases of 33.9\%, 25.5\%, and 53.4\% compared to the three baselines. Notably, Duinocode exhibits higher accuracy than the other two baselines due to its use of partial library knowledge (\ie, library examples). However, none of the baselines achieve full automation, as all require human intervention during the development process.

\begin{table}[t!]
\caption{\textbf{Comparison with baselines.}}
\fontsize{6.3pt}{7.2pt}\selectfont % Custom font size between tiny and scriptsize
\begin{threeparttable}
\scriptsize
\setlength{\tabcolsep}{3.9pt}
\begin{tabular}{l|cc|ccc}
\toprule

\multicolumn{1}{c|}{\textbf{System}} & \multicolumn{2}{c|}{\textbf{Metrics}} & \multicolumn{3}{c}{\textbf{Features\tnote{*}}} \\
\cmidrule(rl){2-3}
\cmidrule(rl){4-6}

& \textbf{Cod. Acc.} & \textbf{Comp. Rate} & \textbf{Lib. Kno.} & \textbf{Debug} & \textbf{Auto.}   \\
\midrule

LLM-Prompt~\cite{englhardt2024exploring} & \timebar{100}{72.3}\% & \timebar{100}{64.6}\% & \Circle & \LEFTcircle & \Circle  \\
Duinocode~\cite{duinocode} & \timebar{100}{82.8}\% & \timebar{100}{68.9}\% & \LEFTcircle & \Circle & \Circle \\
LLM-direct & \timebar{100}{69.5}\% & \timebar{100}{56.4}\% & \Circle & \Circle & \Circle \\
\SystemName & \timebar{100}{95.7}\% & \timebar{100}{86.5}\% & \CIRCLE & \CIRCLE & \CIRCLE\\

\bottomrule
\end{tabular}
\begin{tablenotes}
    \item[*]\CIRCLE \thinspace for applicable, \Circle \thinspace for not applicable, \LEFTcircle \thinspace for partially applicable.
\end{tablenotes}

\end{threeparttable}
\vspace{-0.2in}
\label{tab:baseline}
\end{table}

\subsection{Micro-benchmark Evaluation}

\begin{figure*}
\centering
\subfigure[Impact of library solving.]{
\begin{minipage}[t]{0.24\linewidth}
\centering
\includegraphics[width=0.95\linewidth]{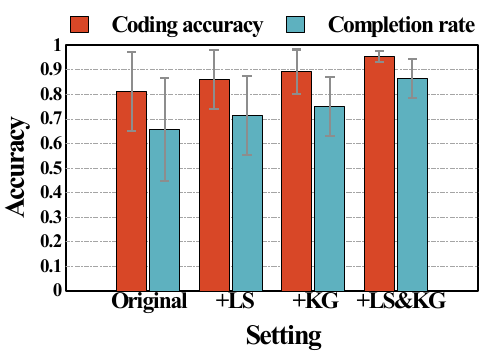}
\label{fig:embed_lib}
\vspace{-4mm}
\end{minipage}%
}%
\subfigure[Impact of library selection number.]{
\begin{minipage}[t]{0.24\linewidth}
\centering
\includegraphics[width=0.95\linewidth]{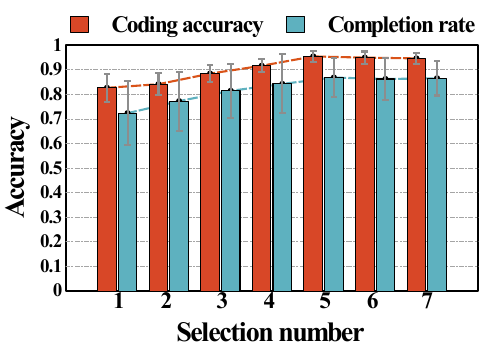}
\label{fig:embed_select}
\vspace{-4mm}
\end{minipage}%
}%
\subfigure[Impact of auto-programming.]{
\begin{minipage}[t]{0.24\linewidth}
\centering
\includegraphics[width=0.95\linewidth]{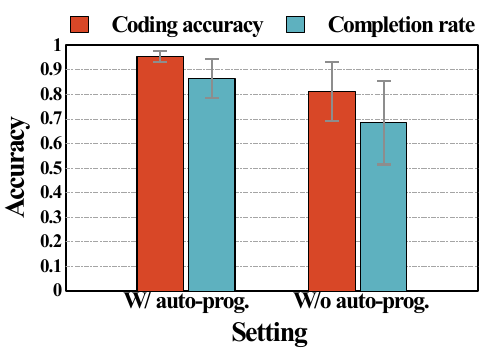}
\label{fig:embed_autodebug}
\vspace{-4mm}
\end{minipage}%
}%
\subfigure[Impact of compile debug trial.]{
\begin{minipage}[t]{0.24\linewidth}
\centering
\includegraphics[width=0.95\linewidth]{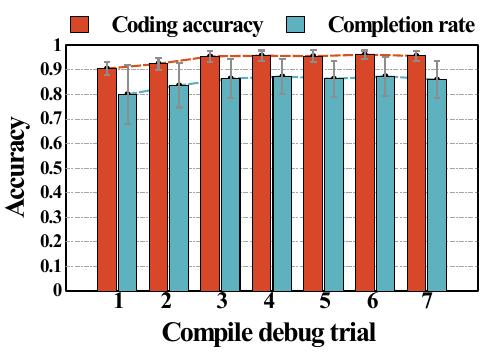}
\label{fig:embed_compile}
\vspace{-4mm}
\end{minipage}%
}%
\\
\vspace{-3mm}
\subfigure[Impact of flash debug trial.]{
\begin{minipage}[t]{0.24\linewidth}
\centering
\includegraphics[width=0.95\linewidth]{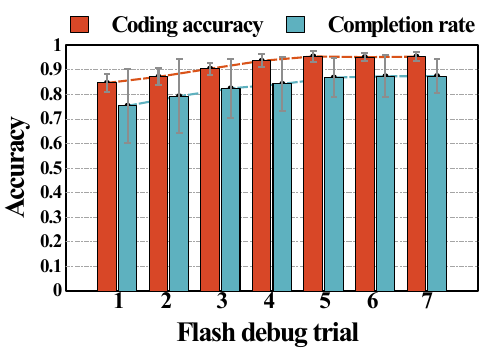}
\label{fig:embed_flash}
\vspace{-4mm}
\end{minipage}%
}%
\subfigure[Impact of development platforms.]{
\begin{minipage}[t]{0.24\linewidth}
\centering
\includegraphics[width=0.95\linewidth]{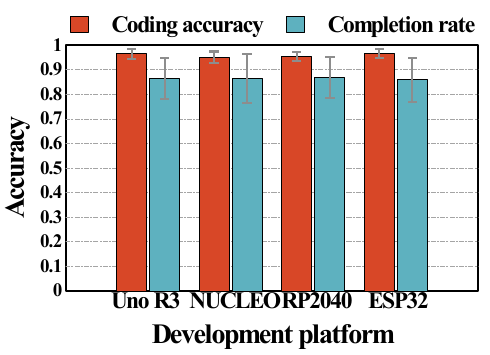}
\label{fig:embed_platform}
\vspace{-4mm}
\end{minipage}%
}%
\subfigure[Impact of library complexity.]{
\begin{minipage}[t]{0.24\linewidth}
\centering
\includegraphics[width=0.95\linewidth]{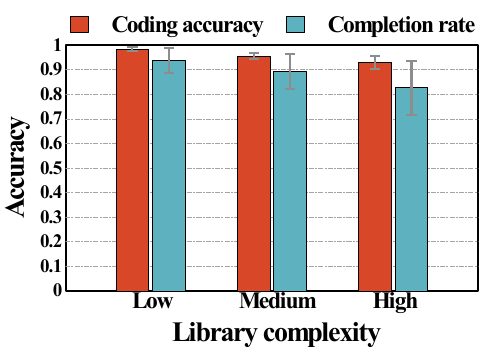}
\label{fig:embed_libcom}
\vspace{-4mm}
\end{minipage}%
}%
\subfigure[Impact of security checker.]{
\begin{minipage}[t]{0.24\linewidth}
\centering
\includegraphics[width=0.95\linewidth]{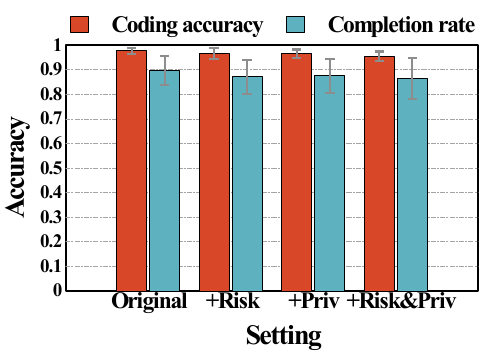}
\label{fig:embed_checker}
\vspace{-4mm}
\end{minipage}%
}%
\vspace{-2mm}
\centering
\caption{\textbf{Experimental results.}}
% \label{fig:deform3}
\vspace{-0.2in}
\end{figure*}

\textbf{Library Solving and Knowledge Generation.}
% with or without(pick the first search result)
% The effectiveness of the library solving is shown in Fig.~\ref{fig:embed_lib}, where we compare the coding accuracy and completion rate utilizing the proposed library solving method against employing the first search result. The results indicate that \SystemName surpasses the method without library solving. Notably, \SystemName achieves an increase in coding accuracy by 10.8\% and completion rate by 21\%. 
% This finding highlight the superiority of \SystemName in library solving, culminating in superior performance.
The effectiveness of the library-solving approach is demonstrated in Fig.~\ref{fig:embed_lib}, where we compare the coding accuracy and completion rate of the proposed method against the use of the first search result. The results clearly show that \SystemName outperforms the approach without library solving, achieving a 10.8\% increase in coding accuracy and a 21\% improvement in the completion rate. 
% This finding underscores the superiority of \SystemName in handling dependency selection challenges, leading to significantly enhanced overall performance.
% \noindent\textbf{Knowledge Generation.}
% with or without(pick the most similar example only)
% We then evaluate the impact of the knowledge generation approach. As demonstrated in Fig.~\ref{fig:embed_kg}, where we compare the coding accuracy and completion rate of the proposed method against the only use of the most similar example to the task. The results clearly show that \SystemName outperforms the approach without library solving, achieving a 10.7\% increase in coding accuracy and a 10.6\% improvement in the completion rate. This finding underscores the superiority of \SystemName in select and use of library knowledge, leading to significantly enhanced overall performance.
We then evaluate the impact of the knowledge generation approach. As shown in Fig.~\ref{fig:embed_lib}, we compare the performance of the proposed method against using only the most similar example for the task. The results demonstrate that \SystemName outperforms the alternative, achieving a 7.1\% increase in coding accuracy and a 15\% improvement in completion rate. This finding highlights the effectiveness of our system in selecting and utilizing library knowledge, resulting in improved overall performance.

% \textbf{Knowledge Memory Pick-up.}
% with or without(pick the first search result)
\noindent\textbf{Library Selection Number.}
% For each component, \SystemName searches for libraries and retrieves the top N libraries for further processing. Here we evaluate the impact of the N.  As shown in Fig.~\ref{fig:embed_select}, performance demonstrated a positive correlation with the number of selection number, an increase from 1 to 3 led to a 7\% improvement in coding accuracy and an 12.6\% increase in completion rate. an increase from 3 to 5 led to a 7.9\% improvement in coding accuracy and an 6.6\% increase in completion rate. This enhancement can be attributed to the sometimes the most fit library did not in the very first search results, we can select it with our Library Selection algorithm. However, when selection number bigger, no further increase observed. This is because the xxxx. To sum up, xxx
\SystemName searches for libraries and retrieves the top N candidates for library selection as introduced in Sec.~\ref{subsubsec:solve_lib}. Here, we evaluate the impact of N on performance. As shown in Fig.~\ref{fig:embed_select}, the performance exhibited a positive correlation with the number of selected libraries. Increasing N from 1 to 3 resulted in a 7\% improvement in coding accuracy and a 12.6\% increase in completion rate. Further increasing N from 3 to 5 led to a 7.9\% improvement in coding accuracy and a 6.6\% increase in completion rate. This enhancement can be attributed to the fact that the most suitable library is not always among the top search results, but our selection algorithm can identify it when more options are considered. However, beyond a certain threshold, no further performance gains were observed. This is because expanding the range too much introduces more irrelevant libraries, which dilutes the effectiveness of the selection. Therefore, we choose 5 as the optimal number.

\noindent\textbf{Auto-Programming.}
% with or without Auto-Debug
% The impact of our proposed Auto-Debugging technique is assessed by comparing performance metrics with and without the application of this method. As depicted in Fig.~\ref{fig:embed_autodebug}, The results clearly demonstrate that \SystemName achieves a 17.6\% increase in coding accuracy and a 26.2\% improvement in completion rate. This finding highlights the effectiveness of \SystemName in xxxx, resulting in improved overall performance.
The impact of our proposed auto-programming technique is evaluated by comparing performance metrics with and without the application of this method. As depicted in Fig.~\ref{fig:embed_autodebug}, the results demonstrate that \SystemName achieves a 17.6\% increase in coding accuracy and a 26.2\% improvement in completion rate. This finding highlights the effectiveness of \SystemName in automating the debugging process, leading to promising performance.

\noindent\textbf{Compile Debug Trials.}
% different N
% In our investigation of the effect of compile debug trials on coding accuracy and completion rate within \SystemName, we observe notable trends as illustrated in Fig.~\ref{fig:embed_compile}. Accuracy exhibits a positive correlation with trial times, where an increase from 1 to 3 trials resulted in a 5.5\% improvement in coding accuracy and 8\% improvement in completion rate, and a further increase to 7 trials is table led to no more improvement. This enhancement can be attributed to the information provided by the compiler for error such as misuse of library, which contributes to the system's ability to correctly recorrect them. Consequently, we recommend maintaining a minimum compile debug trials of 3 to ensure reliable performance.
In our investigation of the effect of compile debug trials on coding accuracy and completion rate within \SystemName, we observed notable trends, as illustrated in Fig.~\ref{fig:embed_compile}. Accuracy demonstrated a positive correlation with the number of trials: an increase from 1 to 3 trials led to a 5.5\% improvement in coding accuracy and an 8\% increase in completion rate. This enhancement can be attributed to the error information provided by the compiler, such as identifying library misuse, which facilitates the system's ability to correct these errors. However, beyond 3 trials, with up to 7 trials tested, no obvious improvements were observed. Therefore, we recommend maintaining a minimum of 3 compile debug trials to ensure reliable performance.

\noindent\textbf{Flash Debug Trials.}
% different N
We then assessed the impact of flash debug trials on coding accuracy and completion rate within \SystemName, revealing significant trends, as depicted in Fig.~\ref{fig:embed_flash}. We observed a positive relationship between accuracy and the number of trials: an increase from 1 to 3 trials resulted in a 6.8\% enhancement in coding accuracy and a 9.4\% rise in completion rate. Furthermore, an increase from 3 to 5 trials led to a 5.5\% improvement in coding accuracy and a 5.3\% increase in completion rate. This improvement can be attributed to the debug information embedded during coding, which enhances the system's error correction capabilities. However, after conducting up to 7 trials, no significant enhancements were noted. Therefore, we recommend using 5 flash debug trials to ensure dependable performance.

\noindent\textbf{Development Platform.}
% cod ACC and comp rate for three boards.
Then, we compared the performance of \SystemName across different development platforms. To ensure a fair comparison, we selected ten modules compatible with all four platforms (\ie, Uno R3, NUCLEO-L4, Nano RP2040, and Nano ESP32
). As shown in Fig.~\ref{fig:embed_platform}, all four platforms achieved over 95\% coding accuracy and over 80\% completion rates. These consistent results can be attributed to \SystemName's efficient automation design, which minimizes platform-specific dependencies and ensures a uniform development experience. This demonstrates that \SystemName delivers consistently high performance across various platforms, indicating its robustness and adaptability to different hardware environments.

% \textbf{Task Complexity.}

\noindent\textbf{Library Complexity.}
% library API numbers
We evaluate the impact of library complexity on \SystemName's performance by categorizing all modules based on the number of library APIs into three levels: low (fewer than 10 APIs), medium (10-20 APIs), and high (more than 20 APIs). As shown in Fig.~\ref{fig:embed_libcom}, \SystemName achieved over 90\% coding accuracy and over 80\% completion rate across all levels of library complexity. These results indicate that \SystemName maintains consistent performance regardless of the complexity of the libraries involved, demonstrating its ability to handle diverse coding challenges.

\noindent\textbf{Security Checker.} 
% We design security checker to protect the risk and privacy issue when xxx. We evaluate \SystemName's performance when add the risk protection and privacy protection information in prompt. As shown in Fig.~\ref{fig:embed_checker}, When risk protection added, a decrease of 1\% in coding accuracy and 2.8\% in completion rate are observed. when privacy protection is added, a decrease of 1.3\% in coding accuracy and 2.3\% in completion rate. When add both, a decrease of 2.3\% in coding accuracy and 3.6\% in completion rate. However, the performance is still high, show its robustness.
We evaluate \SystemName's performance after adding risk protection and privacy protection information to the prompt. As shown in Fig.~\ref{fig:embed_checker}, when risk protection was added, coding accuracy decreased by 1\%, and completion rate decreased by 2.8\%. When privacy protection was added, coding accuracy decreased by 1.3\%, and completion rate decreased by 2.3\%. When both protections were added simultaneously, there was a total decrease of 2.3\% in coding accuracy and 3.6\% in completion rate. However, the system's performance remained high, demonstrating its robustness even with security checker information in prompt.

\begin{figure}
\centering
\subfigure[Setup.]{
\begin{minipage}[t]{0.52\linewidth}
\centering
\includegraphics[width=1.01\linewidth]{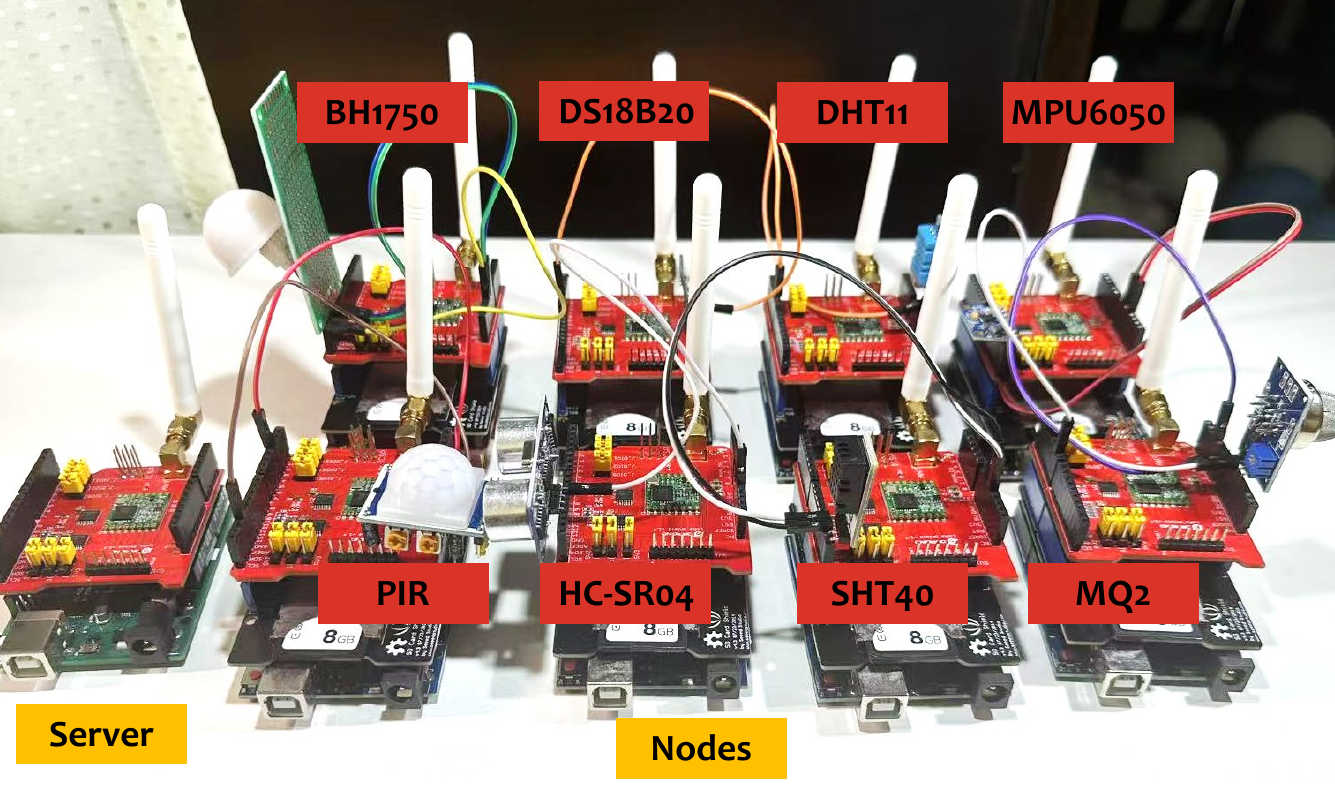}
\label{fig:scale_setup}
\vspace{-4mm}
\end{minipage}%
}
\hspace{-0.1in}
\subfigure[Performance.]{
\begin{minipage}[t]{0.46\linewidth}
\centering
\includegraphics[width=1.02\linewidth]{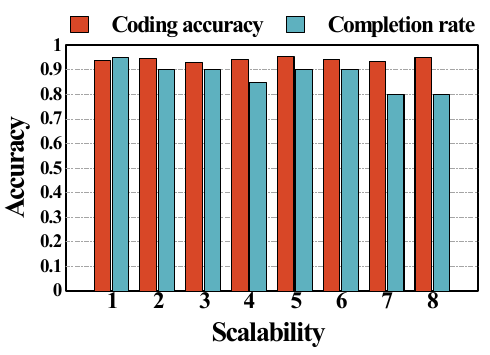}
\label{fig:scale_result}
\vspace{-4mm}
\end{minipage}%
}
\vspace{-2mm}
\centering
\caption{\textbf{Scalability.}}
\label{fig:scale}
\vspace{-0.3in}
\end{figure}

\noindent\textbf{Scalability.} We evaluated the scalability of our system by developing embedded systems of varying scales. Specifically, we tested setups with 1 to 8 LoRa nodes, each transmitting sensor data to a LoRa server. Fig.~\ref{fig:scale_setup} illustrates the LoRa devices used in the evaluation. As shown in Fig.~\ref{fig:scale_result}, the coding accuracy remains stable. While the completion rate decreases as the scale approaches 8, it consistently stays above 80\%, demonstrating the robust scalability of \SystemName.

% three LLM
\noindent\textbf{Overhead.}
% We evaluate the token computation and latency required for \SystemName to finish a develop an embedded system with specific tasks. We select 50 tasks of 10 modules cover sensor, communication, display, and storage from \DatasetName for this evaluation. We use Usage data from the API response to calculate the token consumption of each step. The latency is calculated by using the built-in time calculation function in the debug tool. The token computation and latency of different stages are shown in Tab.~\ref{tab:power}. 
% \SystemName reduces runtime overhead by Selective Memory Pick-up method as mentioned in Sec.~\ref{}, which can reduce the prompt size. Therefore we show \SystemName with Selective Memory Pick-up and without Selective Memory Pick-up for comparison.  \SystemName
% Selective pick API information in the memory
% simplifies the expression of the necessary information, reducing the token
% count by 26.2\% (3.7K on average) and reduce the latency by 11\%. There are two main reason: (i) Reducing token length can significantly decrease
% the on-cloud model’s inference latency. (ii)  it can reduce costs. 
% For example, for GPT-4o,
% the cost can be reduced from \$1.17 to \$0.87 every 10 tasks respectively on average.
We evaluate the token consumption and latency required by \SystemName to develop an embedded system with specific tasks. For this evaluation, we select 50 tasks from 10 modules that cover sensors, communication, display, and storage, sourced from \DatasetName. We calculate token consumption for each step using the usage data from API responses, and latency is measured using the built-in time calculation function in the debug tool. The token consumption and latency for different stages are presented in Tab.~\ref{tab:power}.
\SystemName reduces runtime overhead by employing the selective memory pick-up method, as discussed in Sec.~\ref{subsubsec:selective_memory}, which helps reduce prompt size. Therefore, we compare \SystemName with and without selective memory pick-up. By selectively picking relevant API information, \SystemName simplifies the necessary information, reducing token count by 26.2\% (an average of 3.7K tokens) and latency by 11\%. The benefits of this method include: (i) shorter token lengths decrease the inference latency of the LLM model, and (ii) reduced costs. For instance, with GPT-4o, the cost per 10 tasks decreases from \$1.17 to \$0.87 on average.

\begin{table}[t!]
% \small
\centering
\caption{\textbf{Overhead}.}
\label{tab:power}
\fontsize{6.3pt}{7.2pt}\selectfont % Custom font size between tiny and scriptsize
\scriptsize
\setlength{\tabcolsep}{1.6pt}
\begin{tabular}{lcccc}
\toprule
% \rowcolor{Gray}
\multirow{2}{*}{\diagbox{\textbf{Stg.}}{\textbf{Meth.}}{\textbf{Ind.}}} & \multicolumn{2}{c}{\textbf{Token consumption (K)}} & \multicolumn{2}{c}{\textbf{Latency (s)}} \\
              \cmidrule(lr){2-3}    \cmidrule(lr){4-5} 
                                                                        & \textbf{W/ Sel. Mem.}               & \textbf{W/o Sel. Mem.}                 & \textbf{W/ Sel. Mem.}                   & \textbf{W/o Sel. Mem.}                 \\
                                                                        \hline
\textbf{Lib. Solv.}                                           & -                & -                & 4.9                 & 5.2                   \\
\textbf{Know. Gen.}                                           & 7.925                & 7.893                & 95.7                 & 89.8                   \\
\textbf{Know. Extc. }                                          & 0.224                & 0.232                & 2.3                 & 1.8                   \\
\textbf{Sec. Chk.}                                               & 0.117              & 0.114              & 1.8                  & 1.6                   \\ 
\textbf{Cod. \& Debug.}                                         & 2.173              & 5.874              & 8.9                  & 29.2                   \\ \hline
\textbf{Total}                                                                   & 10.439              & 14.113              & 113.6                  & 127.6          \\
\bottomrule
\end{tabular}
\vspace{-0.1in}
\end{table}

\subsection{Case Study}
In this section, we delve into how \SystemName streamlines the development of intricate embedded systems through two case studies
% ~\footnote{Demo: \href{https://embedgenius3.github.io/\#Demo}{https://embedgenius3.github.io/\#Demo}}
: 1) an environmental monitoring system and 2) a remote control system. 
These case studies were selected for their diverse modules and real-world relevance, such as environmental monitoring for sustainability and remote control for smart home automation.

\subsubsection{Environmental Monitoring.}
%\textbf{Motivation} 
We begin by demonstrating how \SystemName automates the development of an environmental monitoring sensor node. This node combines a sensor for data collection, a transceiver for transmission, and a display for real-time updates. Integrating these devices is time-consuming, requiring hours or days even for experienced engineers, with novices needing even longer.
% This scenario greatly benefits from \SystemName, enabling fully automated embedded system development with human-level accuracy in a fraction of the time.

\begin{figure}
    \centering
\includegraphics[width=0.32\textwidth]{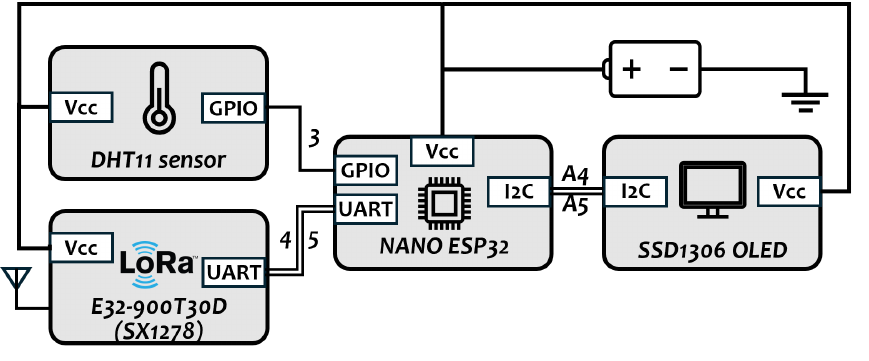}
    \caption{\textbf{Environmental monitoring system architecture.}}
    \label{fig:Case1Stru}
    \vspace{-0.2in}
\end{figure}

\noindent\textbf{Experimental Setup.} 
% The hardware architecture of the climate monitoring system is illustrated in Figure \ref{fig:Case1Stru}. The Nano ESP32 board was chosen as the development platform due to its robust processing capabilities and tiny size. The DHT11 digital temperature sensor interfaces with the Nano ESP32 through pin 3, operating in a single-bus mode to ensure precise temperature readings.
% For data transmission, the system employs the E32-900T30D (SX1278) LoRa module. This module facilitates long-range communication by transmitting data packets to a corresponding module connected to the server. The connection utilizes pins 4 and 5 on the Nano ESP32, configured for serial port transparent transmission mode.
% An OLED I2C 4-pin display module, driven by the SSD1306 IC, is integrated into the system to provide real-time feedback. This display is connected to the Nano ESP32 via the A4 pin (I2C SDA) and A5 pin (I2C SCL), enabling the visualization of system status and operational data.
The hardware architecture of the environmental monitoring system is depicted in Fig.~\ref{fig:Case1Stru}. The Nano ESP32 board was selected as the development platform for its robust processing capabilities and compact size. 
% The DHT11 digital temperature sensor interfaces with the Nano ESP32 through pin 3, operating in a single-bus mode to ensure accurate temperature readings.
% For data transmission, the system utilizes the E32-900T30D (SX1278) LoRa module. This module enables long-range communication by sending data packets to a corresponding module connected to the server. The connection is established using pins 4 and 5 on the Nano ESP32, configured for serial port transparent transmission mode.
% An OLED I2C 4-pin display module, powered by the SSD1306 IC, is integrated into the system to provide real-time feedback. This display is linked to the Nano ESP32 via the A4 pin (I2C SDA) and A5 pin (I2C SCL).
A DHT11 sensor connects via pin 3 for accurate temperature readings, while an E32-900T30D LoRa module on pins 4 and 5 enables long-range data transmission to a server. An SSD1306-based OLED display, connected via A4 (I2C SDA) and A5 (I2C SCL), provides real-time feedback.

\noindent\textbf{Task Description.} \textit{"Develop an environmental monitor system that polls the DHT11 sensor every second to acquire temperature readings, transmits the temperature value to a server using the LoRa module, and updates the OLED display with the current temperature and LoRa transmission status."}

\noindent\textbf{Metadata.} \textit{"Nano ESP32 connected with DHT11 to Pin 3, LoRa module to Pin 4 and 5, and OLED display to Pin A4 and A5."}

% Upon finalizing the hardware setup, the system was autonomously developed and programmed by \SystemName. The primary responsibilities assigned to \SystemName included:
% \begin{itemize}
%     \item Polling the DHT11 sensor every second to acquire temperature readings.
%     \item Transmitting the temperature value to the server using the LoRa module.
%     \item Updating the OLED display with the current temperature and the status of the LoRa transmission after each data exchange cycle.
% \end{itemize}

\begin{figure}
    \centering
\includegraphics[width=0.40\textwidth]{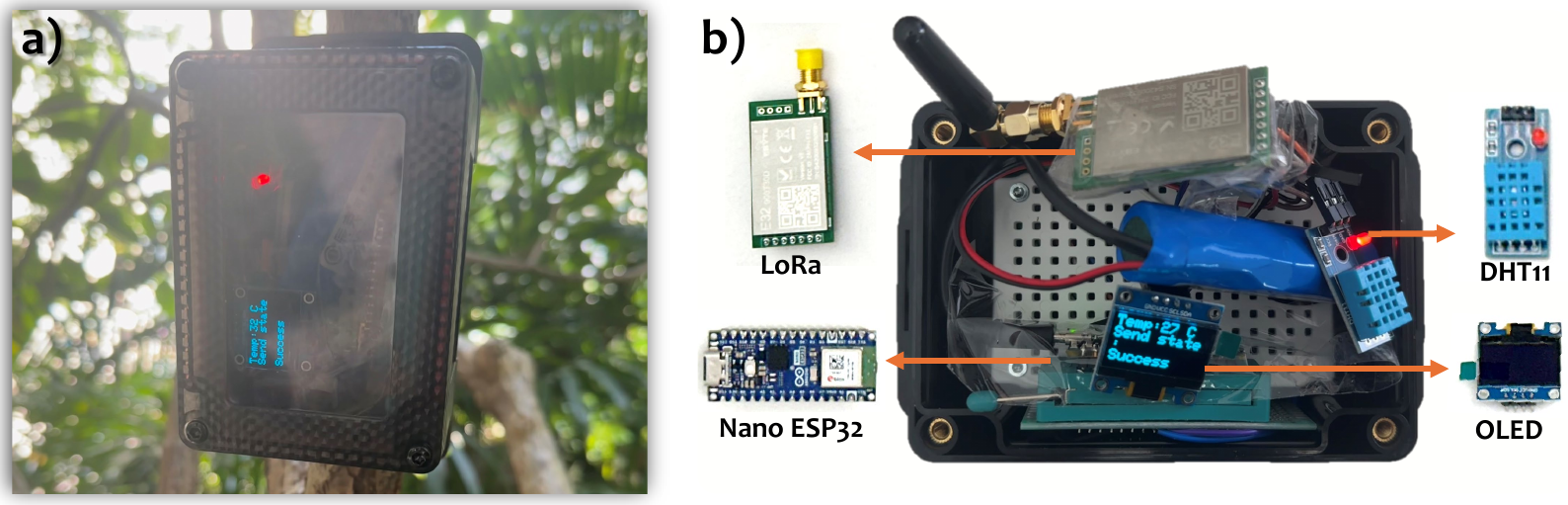}
    \caption{\textbf{Environmental monitoring system deployment.}}
    \label{fig:Case1Env}
    \vspace{-0.2in}
\end{figure}

\begin{figure}
\centering
\subfigure[Performance.]{
\begin{minipage}[t]{0.49\linewidth}
\centering
\includegraphics[width=0.99\linewidth]{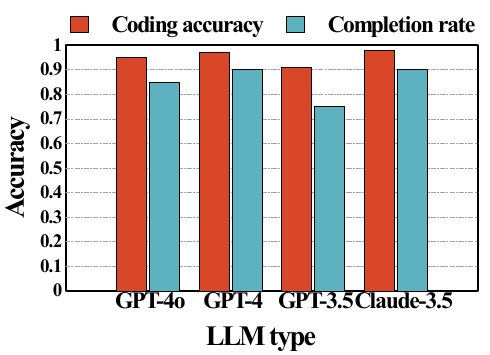}
\label{fig:case1_r1}
\vspace{-4mm}
\end{minipage}%
}
\hspace{-0.1in}
\subfigure[Overhead.]{
\begin{minipage}[t]{0.49\linewidth}
\centering
\includegraphics[width=0.99\linewidth]{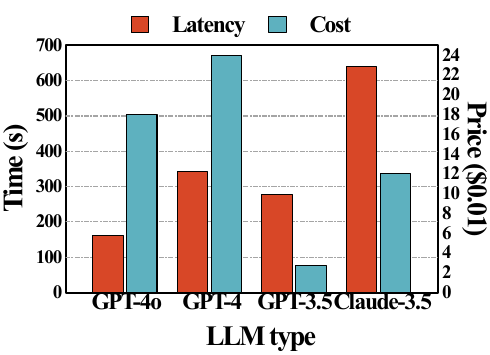}
\label{fig:case1_r2}
\vspace{-4mm}
\end{minipage}%
}
\vspace{-2mm}
\centering
\caption{\textbf{Case study 1 results.}}
\label{fig:case1_result}
\vspace{-0.3in}
\end{figure}

\noindent\textbf{Performance.}
% We evaluate \SystemName to handle the first case study by comparing coding accuracy and completion rate of different LLM. The experiments include 20 times running of this task by \SystemName with each LLM. As shown in Fig.~\ref{fig:case1_r1}, the coding accuracy of GPT4o, GPT4 and CLaude3.5 are higher than 90\%, and the completion rate  are higher than 80\%. The performance of Using GPT3 is lower to about 70\% completion rate. 
As shown in Fig.~\ref{fig:Case1Env}, \SystemName successfully developed the environmental monitoring system, which has been deployed in a forest for continuous environmental monitoring.
We evaluate \SystemName on the first case study by comparing the coding accuracy and completion rate using different LLMs. The experiments involve running this task 20 times using \SystemName with each LLM. As shown in Fig.~\ref{fig:case1_r1}, GPT-4o, GPT-4, and Claude-3.5 achieve coding accuracy above 90\% and completion rates exceeding 80\%. In contrast, GPT-3.5 exhibits a lower performance, with a completion rate of around 70\%. These results demonstrate that newer LLMs, such as GPT-4 and Claude-3.5, are better suited for complex coding tasks.
% , offering higher accuracy and reliability than older models like GPT3.5.

\noindent\textbf{Overhead.} 
% We than evaluate the overhead of \SystemName using different LLMs to finish this task. As shown in Fig.~\ref{fig:case1_r2}, the latency of claude3.5 is over 600s, which is 4 times to GPT4o and 2-3 times to GPT4 and GPT3.5. The price for GPT3.5 is the lowest as it has a very low unit price (\$0.50 / 1M tokens and \$1.50 / 1M tokens). To balance the performance and the overhead, we recommend to use GPT4o, as it has a low latency and a good performance.
We then evaluate the performance impact of \SystemName when using different LLMs to complete this task. As depicted in Fig.~\ref{fig:case1_r2}, Claude3.5 demonstrates the longest latency, exceeding \SI{600}{\second}, which is approximately four times longer than GPT-4o and two to three times longer than GPT-4 and GPT-3.5. Regarding cost efficiency, GPT-3.5 emerges as the most economical choice, thanks to its low unit pricing (\$0.50 per 1M tokens for input and \$1.50 per 1M tokens for output). To strike a balance between performance and overhead, we recommend opting for GPT-4o, as it offers low latency coupled with robust performance.

\subsubsection{Remote Control.}
% Next, we explore \SystemName's capability in developing a remote device control system. Unlike sensor systems, this project requires front-end development, including displaying sensor status on a web page and enabling remote device control. As a result, developers must have a broader skill set, extending beyond embedded development to include proficiency in web programming. 
% Developing such a system requires a coordinated team effort.
We evaluate \SystemName's ability to develop a remote control system, which demands not only embedded development but also front-end skills for web-based sensor status display and remote control. This project requires a broader skill set and coordinated team effort.
% However, our practical experience demonstrates that 
% \SystemName can independently and effectively address these challenges. Its powerful generalization capabilities allow it to generate not only hardware driver and logic relationship code but also HTML for web page implementation. This feature significantly accelerates the development of system prototypes, facilitating rapid innovation and deployment.

%\textbf{Motivation} 

\begin{figure}
    \centering
\includegraphics[width=0.34\textwidth]{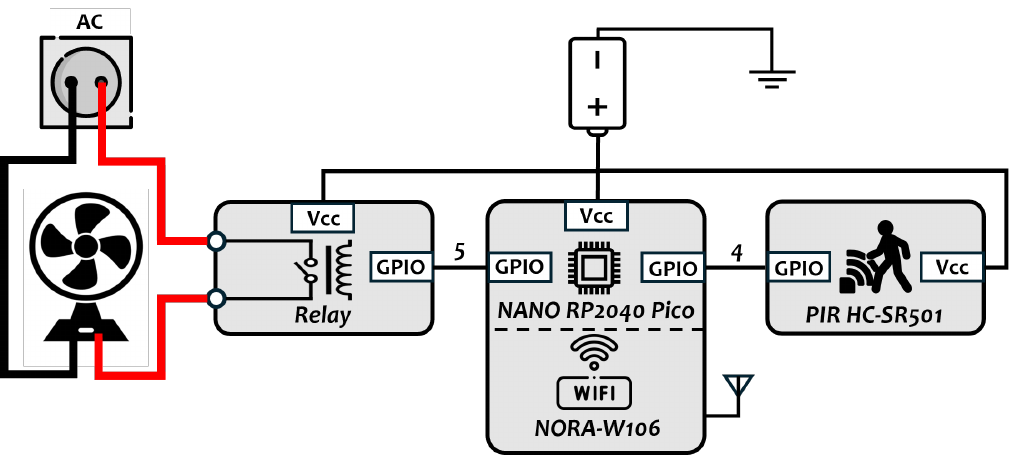}
    \caption{\textbf{Remote control system architecture.}}
    \label{fig:Case2Stru}
    \vspace{-0.1in}
\end{figure}

\noindent\textbf{Experimental Setup.} 
% The hardware architecture of the remote control system is depicted in Fig.~\ref{fig:Case2Stru}. To differentiate from the previous implementation, we employed the Nano RP2040 development plarform, featuring the NORA-W106 Bluetooth Wi-Fi module, as the system's embedded component. 
% The HC-SR501 PIR sensor plays a crucial role in detecting human presence. It transmits signals to pin 4 of the RP2040 PICO, using distinct high and low levels to indicate detection status. Based on the sensor's input, the system controls a relay via pin 5, thereby managing the fan's connection to the AC power supply. This setup enables effective fan control, responding dynamically to occupancy changes.
% The hardware architecture of the remote control system is shown in . Unlike the previous implementation
As shown in Fig.~\ref{fig:Case2Stru}, we utilized the Nano RP2040 development platform with the NORA-W106 Bluetooth Wi-Fi module as the embedded component. The HC-SR501 PIR sensor is responsible for detecting human presence, sending signals to pin 4 of the RP2040 PICO using high and low levels to indicate detection status. Based on the sensor input, the system controls a relay via pin 5 to manage the fan's connection to the AC power supply. 
% This setup allows the fan to respond dynamically to occupancy changes.

\noindent\textbf{Task Description.}
\textit{"Develop a remote control system that polls the PIR sensor every second, controls the fan via a relay, updates a web page with PIR status through the Wi-Fi module (Wi-Fi SSID: WiFiSSID, Password: PASSWORD), and includes a master switch button on the web to control system operation."}

\noindent\textbf{Metadata.} \textit{"Nano RP2040 connected with PIR to Pin 4, relay to Pin 5, and WiFi module on-board."}
% \begin{itemize}
% \item Poll the PIR sensor status every second.
% \item Use a relay to turn the fan on/off based on PIR sensor status.
% \item Update the web page every second to display the PIR status through WIFI module.
% \item Design a button in the web with a master switch to control the entire system operation.
% \end{itemize}

\begin{figure}
    \centering
\includegraphics[width=0.4\textwidth]{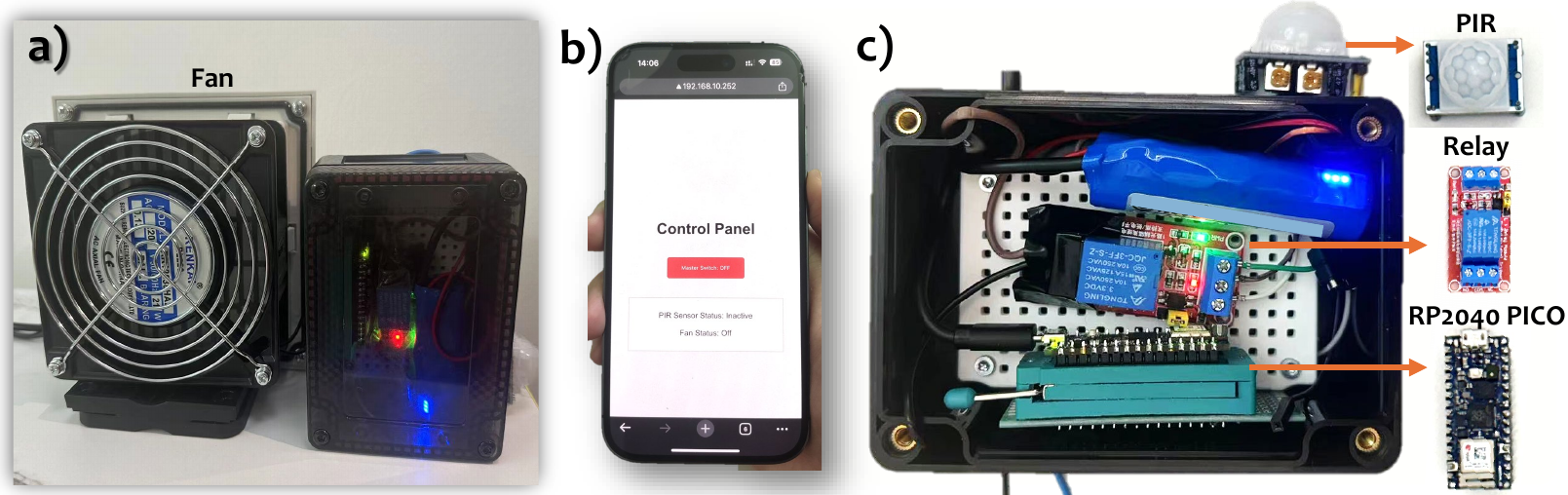}
    \caption{\textbf{Remote control system deployment.}}
    \label{fig:Case2Env}
    \vspace{-0.2in}
\end{figure}

% \noindent\textbf{Performance}
% \begin{figure}
%     \centering
%     \includegraphics[width=0.48\textwidth]{fig/Case2Web.pdf}
%     \caption{\textbf{Webpage designed by \SystemName.}}
%     \label{fig:Case2Web}
% \end{figure}

\begin{figure}[t!]
\centering
\subfigure[Performance.]{
\begin{minipage}[t]{0.49\linewidth}
\centering
\includegraphics[width=0.99\linewidth]{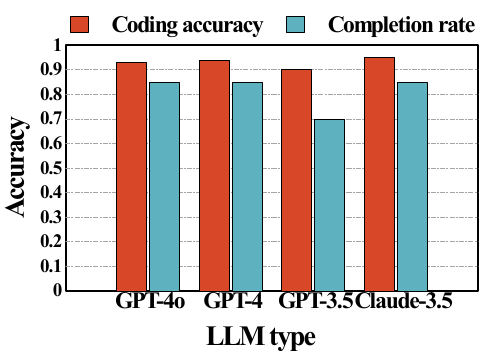}
\label{fig:case2_r1}
\vspace{-4mm}
\end{minipage}%
}
\hspace{-0.1in}
\subfigure[Overhead.]{
\begin{minipage}[t]{0.49\linewidth}
\centering
\includegraphics[width=0.99\linewidth]{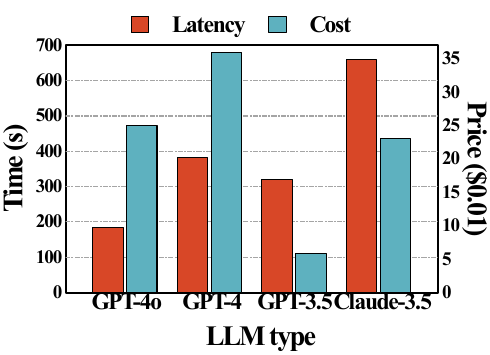}
\label{fig:case2_r2}
\vspace{-4mm}
\end{minipage}%
}
\vspace{-2mm}
\centering
\caption{\textbf{Case study 2 results.}}
\label{fig:case2_result}
\vspace{-0.3in}
\end{figure}

\noindent\textbf{Performance.}
% We evaluate this case study by comparing the performance of \SystemName using different LLMs. As shown in Fig.~\ref{fig:case2_r1}, GPT4o, GPT4, and Claude3.5 achieve completion rates exceeding 80\%, but GPT3.5 is lower than 70\%. This is because this task have some detail need to be handle such as the webpage setup though WIFi, GPT3.5 may not have the ability to make it correct every time. However, they all achieve good performance. For the coding accuracy, four LLM types all above 90\%. These results demonstrate xxx.
As shown in Fig.~\ref{fig:Case2Env}, \SystemName successfully developed the remote fan control system, which has been deployed in a smart home to manage airflow based on occupancy automatically. Notably, in Fig.\ref{fig:Case2Env}(b), the system created a web interface that updates in real-time via Wi-Fi—a capability we had not anticipated \SystemName could achieve.
We evaluate this case study by comparing the performance of \SystemName using different LLMs. As shown in Fig.~\ref{fig:case2_r1}, GPT-4o, GPT-4, and Claude 3.5 achieve completion rates exceeding 80\%, while GPT-3.5 falls below 70\%. This is likely due to the task's complexity, such as handling webpage setup through Wi-Fi, where GPT-3.5 may struggle with consistent accuracy. However, all four LLMs perform well overall, with coding accuracy exceeding 90\%. 
% These results demonstrate \SystemName's effectiveness in generating accurate code across various LLMs, with higher performance observed in more advanced models.

\noindent\textbf{Overhead.} 
% We than evaluate the overhead of \SystemName using different LLMs to finish this task. As shown in Fig.~\ref{fig:case1_r2}, the latency of claude3.5 is over 600s, which is 4 times to GPT4o and 2-3 times to GPT4 and GPT3.5. The price for GPT3.5 is the lowest as it has a very low unit price (\$0.50 / 1M tokens and \$1.50 / 1M tokens). To balance the performance and the overhead, we recommend to use GPT4o, as it has a low latency and a good performance.
We then assess the overhead of \SystemName when utilizing different LLMs to complete this task. As shown in Fig.~\ref{fig:case1_r2}, Claude 3.5 exhibits the longest latency, exceeding \SI{600}{\second}. While GPT-3.5 is theoretically faster, it requires more rounds of debugging, resulting in longer completion times than GPT-4o. In terms of cost, GPT-3.5 remains the most economical choice. Notably, GPT-4o excels in this task, offering low latency and strong performance, striking an ideal balance between speed and accuracy.

\section{Related Work}

\textbf{LLM for Programming.}
Previous research has demonstrated the utilization of language models and AI assistants (\eg, Github Copilot) for generating software, showcasing that modern language models can perform at the level of proficient software developers~\cite{bubeck2023sparks,chen2021evaluating,devlin2017robustfill,gunasekar2023textbooks,jain2022jigsaw,liu2024your,nguyen2022empirical,xu2022systematic,ziegler2022productivity}. 
% Furthermore, several studies have fine-tuned LLMs specifically for code generation in various programming languages, demonstrating enhancements over the base models~\cite{chen2023improving,nijkamp2023codegen2,nijkamp2022codegen}.
Researchers have also devised methods to effectively prompt LLMs for code generation, including self-debugging~\cite{chen2023teaching,zhong2024ldb,shinn2023reflexion}, task decomposition~\cite{kim2024language,yao2023react}, and prompting frameworks~\cite{poesiasynchromesh}.
% The previous studies have assessed LLM-based code generation across various programming languages, such as C/C++ (frequently utilized for embedded software). 
% Previous studies have evaluated LLM-based code generation across languages like C/C++ (commonly used for embedded software).
% Nevertheless, they have limitations in guiding particular hardware devices and incorporating the hardware-specific expertise essential for embedded systems.
Previous studies have evaluated LLM-based code generation across languages like C/C++, but they lack guidance for specific embedded systems and do not cover specific knowledge crucial for developing embedded systems.

\noindent\textbf{LLM for Embedded System.}
Existing research on hardware-platform code generation mainly focuses on tools like IDEs and programming frameworks. However, these studies have not explored the use of LLMs in conjunction with hardware~\cite{ball2019microsoft,brennan2022exploring,devine2018makecode,koopman2005undergraduate,pasricha2022embedded}. Additional systems concentrate on producing hardware-level programs, commonly Hardware Description Language (HDL) code for Field Programmable Gate Arrays (FPGAs)~\cite{guo2008efficient,liu2016automatic,moreira2010automatic}.
% While some tools~\cite{duinocode} exist for LLM-based embedded code generation, 
% and several blog posts and tutorials explore the use of LLMs in embedded development~\cite{puneeth2023chatgpt, dronebot2023chatgpt}, these resources 
% these primarily focus on basic code generation and can not automate the embedded development process.
Furthermore, recent research by Englhardt et al.~\cite{englhardt2024exploring} assesses how LLMs can aid in embedded system development, but the study predominantly focuses on human-in-the-loop tests, lacking automation.
Instead, our work is the first automated embedded system development system designed to handle diverse IoT tasks. 

% \noindent\textbf{LLM in IoT.} Recent studies aim to use sensor data to assist LLM in understanding the physical world~\cite{dai2024advancing,wang2024chattracer,englhardt2024classification,hota2024evaluating,ji2024hargpt,jin2024position,kim2024health,xu2024penetrative,yang2024drhouse,ouyang2024llmsense}.
% For instance, Babel~\cite{dai2024advancing} explores Aligning multiple sensing modalities into one unified representation to bridge LLM for sensing comprehension.
% LLM also for mobile device~\cite{wen2024autodroid,yin2024llm,xu2023llmcad,yi2023edgemoe,zhang2024llamatouch}. For example, AutoDroid~\cite{wen2024autodroid} uses LLMs for mobile device control.
\noindent\textbf{LLM in IoT.} Recent research has increasingly focused on leveraging sensor data to enhance LLMs in understanding the physical world~\cite{dai2024advancing,chen2023rf,wang2024chattracer,englhardt2024classification,hota2024evaluating,ji2024hargpt,jin2024position,kim2024health,xu2024penetrative,yang2024drhouse,ouyang2024llmsense}. For instance, BABEL~\cite{dai2024advancing} explores how aligning multiple sensing modalities into a unified representation can bridge the gap between LLMs and sensory comprehension. Furthermore, LLMs are also being adapted for mobile devices~\cite{wen2024autodroid,yin2024llm,xu2023llmcad,yi2023edgemoe,zhang2024llamatouch}. For example, AutoDroid~\cite{wen2024autodroid} utilizes LLMs to control apps in Android devices with 158 common mobile tasks.

% \section{Discussion}
% \label{sec:discuss}

\section{Conclusion}
\label{sec:conclusion}
This study takes the first step toward automating embedded IoT system development that requires no manual intervention, significantly reducing development time and minimizing errors.
\SystemName introduces a unique set of techniques, such as a library resolution method, a library knowledge generation method, and an auto-programming method.
Extensive evaluation, involving 71 hardware modules, four development platforms, and over 350 IoT tasks, shows that \SystemName achieves an average completion rate of 86.5\%. We will release \SystemName as an open automation tool to facilitate streamlined embedded system development and applications. Future work will focus on extending \SystemName to handle more complex embedded IoT systems.

\bibliographystyle{ACM-Reference-Format}
\bibliography{main}


\begin{thebibliography}{70}

%%% the Web address.

\ifx \showCODEN    \undefined \def \showCODEN     #1{\unskip}     \fi
\ifx \showDOI      \undefined \def \showDOI       #1{#1}\fi
\ifx \showISBNx    \undefined \def \showISBNx     #1{\unskip}     \fi
\ifx \showISBNxiii \undefined \def \showISBNxiii  #1{\unskip}     \fi
\ifx \showISSN     \undefined \def \showISSN      #1{\unskip}     \fi
\ifx \showLCCN     \undefined \def \showLCCN      #1{\unskip}     \fi
\ifx \shownote     \undefined \def \shownote      #1{#1}          \fi
\ifx \showarticletitle \undefined \def \showarticletitle #1{#1}   \fi
\ifx \showURL      \undefined \def \showURL       {\relax}        \fi
% invisible to TeX
\providecommand\bibfield[2]{#2}
\providecommand\bibinfo[2]{#2}
\providecommand\natexlab[1]{#1}
\providecommand\showeprint[2][]{arXiv:#2}

\bibitem[Achiam et~al\mbox{.}(2023)]%
        {achiam2023gpt}
\bibfield{author}{\bibinfo{person}{Josh Achiam}, \bibinfo{person}{Steven Adler}, \bibinfo{person}{Sandhini Agarwal}, \bibinfo{person}{Lama Ahmad}, \bibinfo{person}{Ilge Akkaya}, \bibinfo{person}{Florencia~Leoni Aleman}, \bibinfo{person}{Diogo Almeida}, \bibinfo{person}{Janko Altenschmidt}, \bibinfo{person}{Sam Altman}, \bibinfo{person}{Shyamal Anadkat}, {et~al\mbox{.}}} \bibinfo{year}{2023}\natexlab{}.
\newblock \showarticletitle{Gpt-4 technical report}.
\newblock \bibinfo{journal}{\emph{arXiv preprint arXiv:2303.08774}} (\bibinfo{year}{2023}).
\newblock


\bibitem[Aizawa(2003)]%
        {aizawa2003information}
\bibfield{author}{\bibinfo{person}{Akiko Aizawa}.} \bibinfo{year}{2003}\natexlab{}.
\newblock \showarticletitle{An information-theoretic perspective of tf--idf measures}.
\newblock \bibinfo{journal}{\emph{Elsevier Information Processing \& Management}} (\bibinfo{year}{2003}).
\newblock


\bibitem[{Arduino}(2024a)]%
        {adafruit_ssd1306_library}
\bibfield{author}{\bibinfo{person}{{Arduino}}.} \bibinfo{year}{2024}\natexlab{a}.
\newblock \bibinfo{title}{Adafruit SSD1306 OLED Library}.
\newblock \bibinfo{howpublished}{\url{https://www.arduino.cc/reference/en/libraries/adafruit-ssd1306/}}.
\newblock
\newblock
\shownote{Accessed: 2024-08-30}.


\bibitem[{Arduino}(2024b)]%
        {arduino_library}
\bibfield{author}{\bibinfo{person}{{Arduino}}.} \bibinfo{year}{2024}\natexlab{b}.
\newblock \bibinfo{title}{Arduino Libraries}.
\newblock \bibinfo{howpublished}{\url{https://www.arduino.cc/en/Reference/Libraries}}.
\newblock
\newblock
\shownote{Accessed: 2024-08-22}.


\bibitem[{Arduino}(2024c)]%
        {fastled_library}
\bibfield{author}{\bibinfo{person}{{Arduino}}.} \bibinfo{year}{2024}\natexlab{c}.
\newblock \bibinfo{title}{FastLED Library}.
\newblock \bibinfo{howpublished}{\url{https://www.arduino.cc/reference/en/libraries/fastled/}}.
\newblock
\newblock
\shownote{Accessed: 2024-08-30}.


\bibitem[Ball et~al\mbox{.}(2019)]%
        {ball2019microsoft}
\bibfield{author}{\bibinfo{person}{Thomas Ball}, \bibinfo{person}{Abhijith Chatra}, \bibinfo{person}{Peli de Halleux}, \bibinfo{person}{Steve Hodges}, \bibinfo{person}{Micha{\l} Moskal}, {and} \bibinfo{person}{Jacqueline Russell}.} \bibinfo{year}{2019}\natexlab{}.
\newblock \showarticletitle{Microsoft MakeCode: embedded programming for education, in blocks and TypeScript}. In \bibinfo{booktitle}{\emph{ACM SPLASH}}.
\newblock


\bibitem[Barr and Massa(2006)]%
        {programming2006}
\bibfield{author}{\bibinfo{person}{Michael Barr} {and} \bibinfo{person}{Anthony Massa}.} \bibinfo{year}{2006}\natexlab{}.
\newblock \bibinfo{booktitle}{\emph{Programming Embedded Systems: With C and GNU Development Tools}}.
\newblock \bibinfo{publisher}{O'Reilly Media, Inc.}
\newblock


\bibitem[Bayle(2013)]%
        {bayle2013c}
\bibfield{author}{\bibinfo{person}{Julien Bayle}.} \bibinfo{year}{2013}\natexlab{}.
\newblock \bibinfo{booktitle}{\emph{C programming for Arduino}}.
\newblock \bibinfo{publisher}{Packt Publishing Ltd}.
\newblock


\bibitem[Boehm(1984)]%
        {boehm1984software}
\bibfield{author}{\bibinfo{person}{Barry~W Boehm}.} \bibinfo{year}{1984}\natexlab{}.
\newblock \showarticletitle{Software engineering economics}.
\newblock \bibinfo{journal}{\emph{IEEE TSE}} (\bibinfo{year}{1984}).
\newblock


\bibitem[Brennan and Lesage(2022)]%
        {brennan2022exploring}
\bibfield{author}{\bibinfo{person}{Robert~W Brennan} {and} \bibinfo{person}{Jonathan Lesage}.} \bibinfo{year}{2022}\natexlab{}.
\newblock \showarticletitle{Exploring the implications of OpenAI codex on education for industry 4.0}. In \bibinfo{booktitle}{\emph{Springer SOHOMA}}.
\newblock


\bibitem[Bubeck et~al\mbox{.}(2023)]%
        {bubeck2023sparks}
\bibfield{author}{\bibinfo{person}{S{\'e}bastien Bubeck}, \bibinfo{person}{Varun Chandrasekaran}, \bibinfo{person}{Ronen Eldan}, \bibinfo{person}{Johannes Gehrke}, \bibinfo{person}{Eric Horvitz}, \bibinfo{person}{Ece Kamar}, \bibinfo{person}{Peter Lee}, \bibinfo{person}{Yin~Tat Lee}, \bibinfo{person}{Yuanzhi Li}, \bibinfo{person}{Scott Lundberg}, {et~al\mbox{.}}} \bibinfo{year}{2023}\natexlab{}.
\newblock \showarticletitle{Sparks of artificial general intelligence: Early experiments with gpt-4}.
\newblock \bibinfo{journal}{\emph{arXiv preprint arXiv:2303.12712}} (\bibinfo{year}{2023}).
\newblock


\bibitem[Chen et~al\mbox{.}(2023b)]%
        {chen2023codet}
\bibfield{author}{\bibinfo{person}{Bei Chen}, \bibinfo{person}{Fengji Zhang}, \bibinfo{person}{Anh Nguyen}, \bibinfo{person}{Daoguang Zan}, \bibinfo{person}{Zeqi Lin}, \bibinfo{person}{Jian-Guang Lou}, {and} \bibinfo{person}{Weizhu Chen}.} \bibinfo{year}{2023}\natexlab{b}.
\newblock \showarticletitle{CodeT: Code Generation with Generated Tests}. In \bibinfo{booktitle}{\emph{ICLR}}.
\newblock


\bibitem[Chen et~al\mbox{.}(2021)]%
        {chen2021evaluating}
\bibfield{author}{\bibinfo{person}{Mark Chen}, \bibinfo{person}{Jerry Tworek}, \bibinfo{person}{Heewoo Jun}, \bibinfo{person}{Qiming Yuan}, \bibinfo{person}{Henrique Ponde De~Oliveira Pinto}, \bibinfo{person}{Jared Kaplan}, \bibinfo{person}{Harri Edwards}, \bibinfo{person}{Yuri Burda}, \bibinfo{person}{Nicholas Joseph}, \bibinfo{person}{Greg Brockman}, {et~al\mbox{.}}} \bibinfo{year}{2021}\natexlab{}.
\newblock \showarticletitle{Evaluating large language models trained on code}.
\newblock \bibinfo{journal}{\emph{arXiv preprint arXiv:2107.03374}} (\bibinfo{year}{2021}).
\newblock


\bibitem[Chen et~al\mbox{.}(2023a)]%
        {chen2023teaching}
\bibfield{author}{\bibinfo{person}{Xinyun Chen}, \bibinfo{person}{Maxwell Lin}, \bibinfo{person}{Nathanael Sch{\"a}rli}, {and} \bibinfo{person}{Denny Zhou}.} \bibinfo{year}{2023}\natexlab{a}.
\newblock \showarticletitle{Teaching large language models to self-debug}.
\newblock \bibinfo{journal}{\emph{arXiv preprint arXiv:2304.05128}} (\bibinfo{year}{2023}).
\newblock


\bibitem[Chen and Zhang(2023)]%
        {chen2023rf}
\bibfield{author}{\bibinfo{person}{Xingyu Chen} {and} \bibinfo{person}{Xinyu Zhang}.} \bibinfo{year}{2023}\natexlab{}.
\newblock \showarticletitle{Rf genesis: Zero-shot generalization of mmwave sensing through simulation-based data synthesis and generative diffusion models}. In \bibinfo{booktitle}{\emph{ACM SenSys}}.
\newblock


\bibitem[Chowdhery et~al\mbox{.}(2023)]%
        {chowdhery2023palm}
\bibfield{author}{\bibinfo{person}{Aakanksha Chowdhery}, \bibinfo{person}{Sharan Narang}, \bibinfo{person}{Jacob Devlin}, \bibinfo{person}{Maarten Bosma}, \bibinfo{person}{Gaurav Mishra}, \bibinfo{person}{Adam Roberts}, \bibinfo{person}{Paul Barham}, \bibinfo{person}{Hyung~Won Chung}, \bibinfo{person}{Charles Sutton}, \bibinfo{person}{Sebastian Gehrmann}, {et~al\mbox{.}}} \bibinfo{year}{2023}\natexlab{}.
\newblock \showarticletitle{Palm: Scaling language modeling with pathways}.
\newblock \bibinfo{journal}{\emph{JMLR}} (\bibinfo{year}{2023}).
\newblock


\bibitem[{CJS Robotics}(2024)]%
        {duinocode}
\bibfield{author}{\bibinfo{person}{{CJS Robotics}}.} \bibinfo{year}{2024}\natexlab{}.
\newblock \bibinfo{title}{Duino Code Generator}.
\newblock
\newblock
\urldef\tempurl%
\url{https://www.duinocodegenerator.com/}
\showURL{%
\tempurl}
\newblock
\shownote{Accessed: 2024-08-20}.


\bibitem[Cobbe et~al\mbox{.}(2021)]%
        {cobbe2021training}
\bibfield{author}{\bibinfo{person}{Karl Cobbe}, \bibinfo{person}{Vineet Kosaraju}, \bibinfo{person}{Mohammad Bavarian}, \bibinfo{person}{Mark Chen}, \bibinfo{person}{Heewoo Jun}, \bibinfo{person}{Lukasz Kaiser}, \bibinfo{person}{Matthias Plappert}, \bibinfo{person}{Jerry Tworek}, \bibinfo{person}{Jacob Hilton}, \bibinfo{person}{Reiichiro Nakano}, {et~al\mbox{.}}} \bibinfo{year}{2021}\natexlab{}.
\newblock \showarticletitle{Training verifiers to solve math word problems}.
\newblock \bibinfo{journal}{\emph{arXiv preprint arXiv:2110.14168}} (\bibinfo{year}{2021}).
\newblock


\bibitem[Dai et~al\mbox{.}(2024)]%
        {dai2024advancing}
\bibfield{author}{\bibinfo{person}{Shenghong Dai}, \bibinfo{person}{Shiqi Jiang}, \bibinfo{person}{Yifan Yang}, \bibinfo{person}{Ting Cao}, \bibinfo{person}{Mo Li}, \bibinfo{person}{Suman Banerjee}, {and} \bibinfo{person}{Lili Qiu}.} \bibinfo{year}{2024}\natexlab{}.
\newblock \showarticletitle{Advancing Multi-Modal Sensing Through Expandable Modality Alignment}.
\newblock \bibinfo{journal}{\emph{arXiv preprint arXiv:2407.17777}} (\bibinfo{year}{2024}).
\newblock


\bibitem[Dang et~al\mbox{.}(2022)]%
        {dang2022iotree}
\bibfield{author}{\bibinfo{person}{Tuan Dang}, \bibinfo{person}{Trung Tran}, \bibinfo{person}{Khang Nguyen}, \bibinfo{person}{Tien Pham}, \bibinfo{person}{Nhat Pham}, \bibinfo{person}{Tam Vu}, {and} \bibinfo{person}{Phuc Nguyen}.} \bibinfo{year}{2022}\natexlab{}.
\newblock \showarticletitle{ioTree: a battery-free wearable system with biocompatible sensors for continuous tree health monitoring}. In \bibinfo{booktitle}{\emph{ACM MobiCom}}.
\newblock


\bibitem[Devine et~al\mbox{.}(2018)]%
        {devine2018makecode}
\bibfield{author}{\bibinfo{person}{James Devine}, \bibinfo{person}{Joe Finney}, \bibinfo{person}{Peli De~Halleux}, \bibinfo{person}{Micha{\l} Moskal}, \bibinfo{person}{Thomas Ball}, {and} \bibinfo{person}{Steve Hodges}.} \bibinfo{year}{2018}\natexlab{}.
\newblock \showarticletitle{MakeCode and CODAL: intuitive and efficient embedded systems programming for education}.
\newblock \bibinfo{journal}{\emph{ACM SIGPLAN Notices}} (\bibinfo{year}{2018}).
\newblock


\bibitem[Devlin et~al\mbox{.}(2017)]%
        {devlin2017robustfill}
\bibfield{author}{\bibinfo{person}{Jacob Devlin}, \bibinfo{person}{Jonathan Uesato}, \bibinfo{person}{Surya Bhupatiraju}, \bibinfo{person}{Rishabh Singh}, \bibinfo{person}{Abdel-rahman Mohamed}, {and} \bibinfo{person}{Pushmeet Kohli}.} \bibinfo{year}{2017}\natexlab{}.
\newblock \showarticletitle{Robustfill: Neural program learning under noisy i/o}. In \bibinfo{booktitle}{\emph{PMLR ICML}}.
\newblock


\bibitem[Englhardt et~al\mbox{.}(2024a)]%
        {englhardt2024exploring}
\bibfield{author}{\bibinfo{person}{Zachary Englhardt}, \bibinfo{person}{Richard Li}, \bibinfo{person}{Dilini Nissanka}, \bibinfo{person}{Zhihan Zhang}, \bibinfo{person}{Girish Narayanswamy}, \bibinfo{person}{Joseph Breda}, \bibinfo{person}{Xin Liu}, \bibinfo{person}{Shwetak Patel}, {and} \bibinfo{person}{Vikram Iyer}.} \bibinfo{year}{2024}\natexlab{a}.
\newblock \showarticletitle{Exploring and characterizing large language models for embedded system development and debugging}. In \bibinfo{booktitle}{\emph{ACM CHI EA}}.
\newblock


\bibitem[Englhardt et~al\mbox{.}(2024b)]%
        {englhardt2024classification}
\bibfield{author}{\bibinfo{person}{Zachary Englhardt}, \bibinfo{person}{Chengqian Ma}, \bibinfo{person}{Margaret~E Morris}, \bibinfo{person}{Chun-Cheng Chang}, \bibinfo{person}{Xuhai"~Orson" Xu}, \bibinfo{person}{Lianhui Qin}, \bibinfo{person}{Daniel McDuff}, \bibinfo{person}{Xin Liu}, \bibinfo{person}{Shwetak Patel}, {and} \bibinfo{person}{Vikram Iyer}.} \bibinfo{year}{2024}\natexlab{b}.
\newblock \showarticletitle{From Classification to Clinical Insights: Towards Analyzing and Reasoning About Mobile and Behavioral Health Data With Large Language Models}.
\newblock \bibinfo{journal}{\emph{ACM IMWUT}} (\bibinfo{year}{2024}).
\newblock


\bibitem[Evans(2011)]%
        {evans2011beginning}
\bibfield{author}{\bibinfo{person}{Brian Evans}.} \bibinfo{year}{2011}\natexlab{}.
\newblock \bibinfo{booktitle}{\emph{Beginning arduino programming}}.
\newblock \bibinfo{publisher}{Apress}.
\newblock


\bibitem[Gunasekar et~al\mbox{.}(2023)]%
        {gunasekar2023textbooks}
\bibfield{author}{\bibinfo{person}{Suriya Gunasekar}, \bibinfo{person}{Yi Zhang}, \bibinfo{person}{Jyoti Aneja}, \bibinfo{person}{Caio C{\'e}sar~Teodoro Mendes}, \bibinfo{person}{Allie Del~Giorno}, \bibinfo{person}{Sivakanth Gopi}, \bibinfo{person}{Mojan Javaheripi}, \bibinfo{person}{Piero Kauffmann}, \bibinfo{person}{Gustavo de Rosa}, \bibinfo{person}{Olli Saarikivi}, {et~al\mbox{.}}} \bibinfo{year}{2023}\natexlab{}.
\newblock \showarticletitle{Textbooks are all you need}.
\newblock \bibinfo{journal}{\emph{arXiv preprint arXiv:2306.11644}} (\bibinfo{year}{2023}).
\newblock


\bibitem[Guo et~al\mbox{.}(2008)]%
        {guo2008efficient}
\bibfield{author}{\bibinfo{person}{Zhi Guo}, \bibinfo{person}{Walid Najjar}, {and} \bibinfo{person}{Betul Buyukkurt}.} \bibinfo{year}{2008}\natexlab{}.
\newblock \showarticletitle{Efficient hardware code generation for FPGAs}.
\newblock \bibinfo{journal}{\emph{ACM TACO}} (\bibinfo{year}{2008}).
\newblock


\bibitem[Hota et~al\mbox{.}(2024)]%
        {hota2024evaluating}
\bibfield{author}{\bibinfo{person}{Aritra Hota}, \bibinfo{person}{Soumyajit Chatterjee}, {and} \bibinfo{person}{Sandip Chakraborty}.} \bibinfo{year}{2024}\natexlab{}.
\newblock \showarticletitle{Evaluating Large Language Models as Virtual Annotators for Time-series Physical Sensing Data}.
\newblock \bibinfo{journal}{\emph{arXiv preprint arXiv:2403.01133}} (\bibinfo{year}{2024}).
\newblock


\bibitem[Jain et~al\mbox{.}(2022)]%
        {jain2022jigsaw}
\bibfield{author}{\bibinfo{person}{Naman Jain}, \bibinfo{person}{Skanda Vaidyanath}, \bibinfo{person}{Arun Iyer}, \bibinfo{person}{Nagarajan Natarajan}, \bibinfo{person}{Suresh Parthasarathy}, \bibinfo{person}{Sriram Rajamani}, {and} \bibinfo{person}{Rahul Sharma}.} \bibinfo{year}{2022}\natexlab{}.
\newblock \showarticletitle{Jigsaw: Large language models meet program synthesis}. In \bibinfo{booktitle}{\emph{ICSE}}.
\newblock


\bibitem[Ji et~al\mbox{.}(2024)]%
        {ji2024hargpt}
\bibfield{author}{\bibinfo{person}{Sijie Ji}, \bibinfo{person}{Xinzhe Zheng}, {and} \bibinfo{person}{Chenshu Wu}.} \bibinfo{year}{2024}\natexlab{}.
\newblock \showarticletitle{HARGPT: Are LLMs Zero-Shot Human Activity Recognizers?}
\newblock \bibinfo{journal}{\emph{arXiv preprint arXiv:2403.02727}} (\bibinfo{year}{2024}).
\newblock


\bibitem[Jim{\'e}nez et~al\mbox{.}(2013)]%
        {jimenez2013introduction}
\bibfield{author}{\bibinfo{person}{Manuel Jim{\'e}nez}, \bibinfo{person}{Rogelio Palomera}, {and} \bibinfo{person}{Isidoro Couvertier}.} \bibinfo{year}{2013}\natexlab{}.
\newblock \bibinfo{booktitle}{\emph{Introduction to embedded systems}}.
\newblock \bibinfo{publisher}{Springer}.
\newblock


\bibitem[Jin et~al\mbox{.}(2024)]%
        {jin2024position}
\bibfield{author}{\bibinfo{person}{Ming Jin}, \bibinfo{person}{Yifan Zhang}, \bibinfo{person}{Wei Chen}, \bibinfo{person}{Kexin Zhang}, \bibinfo{person}{Yuxuan Liang}, \bibinfo{person}{Bin Yang}, \bibinfo{person}{Jindong Wang}, \bibinfo{person}{Shirui Pan}, {and} \bibinfo{person}{Qingsong Wen}.} \bibinfo{year}{2024}\natexlab{}.
\newblock \showarticletitle{Position: What Can Large Language Models Tell Us about Time Series Analysis}. In \bibinfo{booktitle}{\emph{ICML}}.
\newblock


\bibitem[Kim et~al\mbox{.}(2024a)]%
        {kim2024language}
\bibfield{author}{\bibinfo{person}{Geunwoo Kim}, \bibinfo{person}{Pierre Baldi}, {and} \bibinfo{person}{Stephen McAleer}.} \bibinfo{year}{2024}\natexlab{a}.
\newblock \showarticletitle{Language models can solve computer tasks}.
\newblock \bibinfo{journal}{\emph{NeurIPS}} (\bibinfo{year}{2024}).
\newblock


\bibitem[Kim et~al\mbox{.}(2024b)]%
        {kim2024health}
\bibfield{author}{\bibinfo{person}{Yubin Kim}, \bibinfo{person}{Xuhai Xu}, \bibinfo{person}{Daniel McDuff}, \bibinfo{person}{Cynthia Breazeal}, {and} \bibinfo{person}{Hae~Won Park}.} \bibinfo{year}{2024}\natexlab{b}.
\newblock \showarticletitle{Health-llm: Large language models for health prediction via wearable sensor data}.
\newblock \bibinfo{journal}{\emph{arXiv preprint arXiv:2401.06866}} (\bibinfo{year}{2024}).
\newblock


\bibitem[Koopman et~al\mbox{.}(2005)]%
        {koopman2005undergraduate}
\bibfield{author}{\bibinfo{person}{Philip Koopman}, \bibinfo{person}{Howie Choset}, \bibinfo{person}{Rajeev Gandhi}, \bibinfo{person}{Bruce Krogh}, \bibinfo{person}{Diana Marculescu}, \bibinfo{person}{Priya Narasimhan}, \bibinfo{person}{Joann~M Paul}, \bibinfo{person}{Ragunathan Rajkumar}, \bibinfo{person}{Daniel Siewiorek}, \bibinfo{person}{Asim Smailagic}, {et~al\mbox{.}}} \bibinfo{year}{2005}\natexlab{}.
\newblock \showarticletitle{Undergraduate embedded system education at Carnegie Mellon}.
\newblock \bibinfo{journal}{\emph{ACM TECS}} (\bibinfo{year}{2005}).
\newblock


\bibitem[Liller et~al\mbox{.}(2023)]%
        {liller2023towards}
\bibfield{author}{\bibinfo{person}{Jackson Liller}, \bibinfo{person}{Trung Tran}, {and} \bibinfo{person}{Phuc"~VP" Nguyen}.} \bibinfo{year}{2023}\natexlab{}.
\newblock \showarticletitle{Towards the Internet of Living Trees for Precision Agriculture}.
\newblock \bibinfo{journal}{\emph{GetMobile: Mobile Computing and Communications}} (\bibinfo{year}{2023}).
\newblock


\bibitem[Liu et~al\mbox{.}(2024)]%
        {liu2024your}
\bibfield{author}{\bibinfo{person}{Jiawei Liu}, \bibinfo{person}{Chunqiu~Steven Xia}, \bibinfo{person}{Yuyao Wang}, {and} \bibinfo{person}{Lingming Zhang}.} \bibinfo{year}{2024}\natexlab{}.
\newblock \showarticletitle{Is your code generated by chatgpt really correct? rigorous evaluation of large language models for code generation}.
\newblock \bibinfo{journal}{\emph{NeurIPS}} (\bibinfo{year}{2024}).
\newblock


\bibitem[Liu et~al\mbox{.}(2016)]%
        {liu2016automatic}
\bibfield{author}{\bibinfo{person}{Zhiqiang Liu}, \bibinfo{person}{Yong Dou}, \bibinfo{person}{Jingfei Jiang}, {and} \bibinfo{person}{Jinwei Xu}.} \bibinfo{year}{2016}\natexlab{}.
\newblock \showarticletitle{Automatic code generation of convolutional neural networks in FPGA implementation}. In \bibinfo{booktitle}{\emph{IEEE FPT}}.
\newblock


\bibitem[Moreira et~al\mbox{.}(2010)]%
        {moreira2010automatic}
\bibfield{author}{\bibinfo{person}{Tomas~G Moreira}, \bibinfo{person}{Marco~A Wehrmeister}, \bibinfo{person}{Carlos~E Pereira}, \bibinfo{person}{Jean-Francois Petin}, {and} \bibinfo{person}{Eric Levrat}.} \bibinfo{year}{2010}\natexlab{}.
\newblock \showarticletitle{Automatic code generation for embedded systems: From UML specifications to VHDL code}. In \bibinfo{booktitle}{\emph{IEEE INDIN}}.
\newblock


\bibitem[Neelakandan et~al\mbox{.}(2021)]%
        {neelakandan2021iot}
\bibfield{author}{\bibinfo{person}{SBMATSDVBBI Neelakandan}, \bibinfo{person}{MA Berlin}, \bibinfo{person}{Sandesh Tripathi}, \bibinfo{person}{V~Brindha Devi}, \bibinfo{person}{Indu Bhardwaj}, {and} \bibinfo{person}{N Arulkumar}.} \bibinfo{year}{2021}\natexlab{}.
\newblock \showarticletitle{IoT-based traffic prediction and traffic signal control system for smart city}.
\newblock \bibinfo{journal}{\emph{Soft Computing}} (\bibinfo{year}{2021}).
\newblock


\bibitem[Nguyen and Nadi(2022)]%
        {nguyen2022empirical}
\bibfield{author}{\bibinfo{person}{Nhan Nguyen} {and} \bibinfo{person}{Sarah Nadi}.} \bibinfo{year}{2022}\natexlab{}.
\newblock \showarticletitle{An empirical evaluation of GitHub copilot's code suggestions}. In \bibinfo{booktitle}{\emph{MSR}}.
\newblock


\bibitem[Ni et~al\mbox{.}(2023)]%
        {ni2023lever}
\bibfield{author}{\bibinfo{person}{Ansong Ni}, \bibinfo{person}{Srini Iyer}, \bibinfo{person}{Dragomir Radev}, \bibinfo{person}{Ves Stoyanov}, \bibinfo{person}{Wen-tau Yih}, \bibinfo{person}{Sida~I Wang}, {and} \bibinfo{person}{Xi~Victoria Lin}.} \bibinfo{year}{2023}\natexlab{}.
\newblock \showarticletitle{Lever: Learning to verify language-to-code generation with execution}. In \bibinfo{booktitle}{\emph{ICML}}.
\newblock


\bibitem[{OpenAI}(2022)]%
        {openai2022chatgpt}
\bibfield{author}{\bibinfo{person}{{OpenAI}}.} \bibinfo{year}{2022}\natexlab{}.
\newblock \bibinfo{title}{ChatGPT}.
\newblock \bibinfo{howpublished}{\url{https://openai.com/blog/chatgpt/}}.
\newblock
\newblock
\shownote{Accessed: 2024-08-22}.


\bibitem[Ouyang and Srivastava(2024)]%
        {ouyang2024llmsense}
\bibfield{author}{\bibinfo{person}{Xiaomin Ouyang} {and} \bibinfo{person}{Mani Srivastava}.} \bibinfo{year}{2024}\natexlab{}.
\newblock \showarticletitle{LLMSense: Harnessing LLMs for High-level Reasoning Over Spatiotemporal Sensor Traces}.
\newblock \bibinfo{journal}{\emph{arXiv preprint arXiv:2403.19857}} (\bibinfo{year}{2024}).
\newblock


\bibitem[Pasricha(2022)]%
        {pasricha2022embedded}
\bibfield{author}{\bibinfo{person}{Sudeep Pasricha}.} \bibinfo{year}{2022}\natexlab{}.
\newblock \showarticletitle{Embedded systems education in the 2020s: Challenges, reflections, and future directions}. In \bibinfo{booktitle}{\emph{GLSVLSI}}.
\newblock


\bibitem[Pham et~al\mbox{.}(2022)]%
        {pham2022pros}
\bibfield{author}{\bibinfo{person}{Nhat Pham}, \bibinfo{person}{Hong Jia}, \bibinfo{person}{Minh Tran}, \bibinfo{person}{Tuan Dinh}, \bibinfo{person}{Nam Bui}, \bibinfo{person}{Young Kwon}, \bibinfo{person}{Dong Ma}, \bibinfo{person}{Phuc Nguyen}, \bibinfo{person}{Cecilia Mascolo}, {and} \bibinfo{person}{Tam Vu}.} \bibinfo{year}{2022}\natexlab{}.
\newblock \showarticletitle{PROS: an efficient pattern-driven compressive sensing framework for low-power biopotential-based wearables with on-chip intelligence}. In \bibinfo{booktitle}{\emph{ACM MobiCom}}.
\newblock


\bibitem[Poesia et~al\mbox{.}(2022)]%
        {poesiasynchromesh}
\bibfield{author}{\bibinfo{person}{Gabriel Poesia}, \bibinfo{person}{Alex Polozov}, \bibinfo{person}{Vu Le}, \bibinfo{person}{Ashish Tiwari}, \bibinfo{person}{Gustavo Soares}, \bibinfo{person}{Christopher Meek}, {and} \bibinfo{person}{Sumit Gulwani}.} \bibinfo{year}{2022}\natexlab{}.
\newblock \showarticletitle{Synchromesh: Reliable Code Generation from Pre-trained Language Models}. In \bibinfo{booktitle}{\emph{ICLR}}.
\newblock


\bibitem[{Precedence Research}(2024)]%
        {precedenceresearch2024}
\bibfield{author}{\bibinfo{person}{{Precedence Research}}.} \bibinfo{year}{2024}\natexlab{}.
\newblock \bibinfo{title}{Embedded Systems Market Size, Share, Trends, Report 2023 to 2032}.
\newblock \bibinfo{howpublished}{\url{https://www.precedenceresearch.com/embedded-systems-market}}.
\newblock
\newblock
\shownote{Accessed: 2024-09-03}.


\bibitem[Revathy et~al\mbox{.}(2017)]%
        {revathy2017automation}
\bibfield{author}{\bibinfo{person}{M Revathy}, \bibinfo{person}{S Ramya}, \bibinfo{person}{R Sathiyavathi}, \bibinfo{person}{B Bharathi}, {and} \bibinfo{person}{V~Maria Anu}.} \bibinfo{year}{2017}\natexlab{}.
\newblock \showarticletitle{Automation of street light for smart city}. In \bibinfo{booktitle}{\emph{IEEE ICCSP}}.
\newblock


\bibitem[{SaM Solutions}(2024)]%
        {SaMSolutions2024}
\bibfield{author}{\bibinfo{person}{{SaM Solutions}}.} \bibinfo{year}{2024}\natexlab{}.
\newblock \bibinfo{title}{All You Need to Know About Embedded System Programming}.
\newblock
\newblock
\urldef\tempurl%
\url{https://www.sam-solutions.com/blog/all-you-need-to-know-about-embedded-system-programming/}
\showURL{%
\tempurl}
\newblock
\shownote{Accessed: 2024-07-29}.


\bibitem[Santos et~al\mbox{.}(2018)]%
        {santos2018portolivinglab}
\bibfield{author}{\bibinfo{person}{Pedro~M Santos}, \bibinfo{person}{Jo{\~a}o~GP Rodrigues}, \bibinfo{person}{Susana~B Cruz}, \bibinfo{person}{Tiago Louren{\c{c}}o}, \bibinfo{person}{Pedro~M d’Orey}, \bibinfo{person}{Yunior Luis}, \bibinfo{person}{Cec{\'\i}lia Rocha}, \bibinfo{person}{Sofia Sousa}, \bibinfo{person}{S{\'e}rgio Cris{\'o}stomo}, \bibinfo{person}{Cristina Queir{\'o}s}, {et~al\mbox{.}}} \bibinfo{year}{2018}\natexlab{}.
\newblock \showarticletitle{PortoLivingLab: An IoT-based sensing platform for smart cities}.
\newblock \bibinfo{journal}{\emph{IEEE IoTJ}} (\bibinfo{year}{2018}).
\newblock


\bibitem[Shi et~al\mbox{.}(2024)]%
        {shi2024soar}
\bibfield{author}{\bibinfo{person}{Shuyao Shi}, \bibinfo{person}{Neiwen Ling}, \bibinfo{person}{Zhehao Jiang}, \bibinfo{person}{Xuan Huang}, \bibinfo{person}{Yuze He}, \bibinfo{person}{Xiaoguang Zhao}, \bibinfo{person}{Bufang Yang}, \bibinfo{person}{Chen Bian}, \bibinfo{person}{Jingfei Xia}, \bibinfo{person}{Zhenyu Yan}, {et~al\mbox{.}}} \bibinfo{year}{2024}\natexlab{}.
\newblock \showarticletitle{Soar: Design and Deployment of A Smart Roadside Infrastructure System for Autonomous Driving}. In \bibinfo{booktitle}{\emph{ACM MobiCom}}.
\newblock


\bibitem[Shinn et~al\mbox{.}(2023)]%
        {shinn2023reflexion}
\bibfield{author}{\bibinfo{person}{Noah Shinn}, \bibinfo{person}{Beck Labash}, {and} \bibinfo{person}{Ashwin Gopinath}.} \bibinfo{year}{2023}\natexlab{}.
\newblock \showarticletitle{Reflexion: an autonomous agent with dynamic memory and self-reflection}.
\newblock \bibinfo{journal}{\emph{arXiv preprint arXiv:2303.11366}} (\bibinfo{year}{2023}).
\newblock


\bibitem[Touvron et~al\mbox{.}(2023)]%
        {touvron2023llama}
\bibfield{author}{\bibinfo{person}{Hugo Touvron}, \bibinfo{person}{Thibaut Lavril}, \bibinfo{person}{Gautier Izacard}, \bibinfo{person}{Xavier Martinet}, \bibinfo{person}{Marie-Anne Lachaux}, \bibinfo{person}{Timoth{\'e}e Lacroix}, \bibinfo{person}{Baptiste Rozi{\`e}re}, \bibinfo{person}{Naman Goyal}, \bibinfo{person}{Eric Hambro}, \bibinfo{person}{Faisal Azhar}, {et~al\mbox{.}}} \bibinfo{year}{2023}\natexlab{}.
\newblock \showarticletitle{Llama: Open and efficient foundation language models}.
\newblock \bibinfo{journal}{\emph{arXiv preprint arXiv:2302.13971}} (\bibinfo{year}{2023}).
\newblock


\bibitem[Truong et~al\mbox{.}(2020)]%
        {truong2020painometry}
\bibfield{author}{\bibinfo{person}{Hoang Truong}, \bibinfo{person}{Nam Bui}, \bibinfo{person}{Zohreh Raghebi}, \bibinfo{person}{Marta Ceko}, \bibinfo{person}{Nhat Pham}, \bibinfo{person}{Phuc Nguyen}, \bibinfo{person}{Anh Nguyen}, \bibinfo{person}{Taeho Kim}, \bibinfo{person}{Katrina Siegfried}, \bibinfo{person}{Evan Stene}, {et~al\mbox{.}}} \bibinfo{year}{2020}\natexlab{}.
\newblock \showarticletitle{Painometry: wearable and objective quantification system for acute postoperative pain}. In \bibinfo{booktitle}{\emph{ACM MobiSys}}.
\newblock


\bibitem[Vasisht et~al\mbox{.}(2017)]%
        {vasisht2017farmbeats}
\bibfield{author}{\bibinfo{person}{Deepak Vasisht}, \bibinfo{person}{Zerina Kapetanovic}, \bibinfo{person}{Jongho Won}, \bibinfo{person}{Xinxin Jin}, \bibinfo{person}{Ranveer Chandra}, \bibinfo{person}{Sudipta Sinha}, \bibinfo{person}{Ashish Kapoor}, \bibinfo{person}{Madhusudhan Sudarshan}, {and} \bibinfo{person}{Sean Stratman}.} \bibinfo{year}{2017}\natexlab{}.
\newblock \showarticletitle{FarmBeats: an IoT platform for Data-Driven agriculture}. In \bibinfo{booktitle}{\emph{USENIX NSDI}}.
\newblock


\bibitem[Vaswani et~al\mbox{.}(2017)]%
        {NIPS2017_3f5ee243}
\bibfield{author}{\bibinfo{person}{Ashish Vaswani}, \bibinfo{person}{Noam Shazeer}, \bibinfo{person}{Niki Parmar}, \bibinfo{person}{Jakob Uszkoreit}, \bibinfo{person}{Llion Jones}, \bibinfo{person}{Aidan~N Gomez}, \bibinfo{person}{\L~ukasz Kaiser}, {and} \bibinfo{person}{Illia Polosukhin}.} \bibinfo{year}{2017}\natexlab{}.
\newblock \showarticletitle{Attention is All you Need}. In \bibinfo{booktitle}{\emph{NeurIPS}}.
\newblock


\bibitem[Wang et~al\mbox{.}(2024)]%
        {wang2024chattracer}
\bibfield{author}{\bibinfo{person}{Qijun Wang}, \bibinfo{person}{Shichen Zhang}, \bibinfo{person}{Kunzhe Song}, {and} \bibinfo{person}{Huacheng Zeng}.} \bibinfo{year}{2024}\natexlab{}.
\newblock \showarticletitle{ChatTracer: Large Language Model Powered Real-time Bluetooth Device Tracking System}.
\newblock \bibinfo{journal}{\emph{arXiv preprint arXiv:2403.19833}} (\bibinfo{year}{2024}).
\newblock


\bibitem[Wei et~al\mbox{.}(2022)]%
        {wei2022chain}
\bibfield{author}{\bibinfo{person}{Jason Wei}, \bibinfo{person}{Xuezhi Wang}, \bibinfo{person}{Dale Schuurmans}, \bibinfo{person}{Maarten Bosma}, \bibinfo{person}{Fei Xia}, \bibinfo{person}{Ed Chi}, \bibinfo{person}{Quoc~V Le}, \bibinfo{person}{Denny Zhou}, {et~al\mbox{.}}} \bibinfo{year}{2022}\natexlab{}.
\newblock \showarticletitle{Chain-of-thought prompting elicits reasoning in large language models}.
\newblock \bibinfo{journal}{\emph{NeurIPS}} (\bibinfo{year}{2022}).
\newblock


\bibitem[Wen et~al\mbox{.}(2024)]%
        {wen2024autodroid}
\bibfield{author}{\bibinfo{person}{Hao Wen}, \bibinfo{person}{Yuanchun Li}, \bibinfo{person}{Guohong Liu}, \bibinfo{person}{Shanhui Zhao}, \bibinfo{person}{Tao Yu}, \bibinfo{person}{Toby Jia-Jun Li}, \bibinfo{person}{Shiqi Jiang}, \bibinfo{person}{Yunhao Liu}, \bibinfo{person}{Yaqin Zhang}, {and} \bibinfo{person}{Yunxin Liu}.} \bibinfo{year}{2024}\natexlab{}.
\newblock \showarticletitle{Autodroid: Llm-powered task automation in android}. In \bibinfo{booktitle}{\emph{ACM MobiCom}}.
\newblock


\bibitem[Xu et~al\mbox{.}(2023)]%
        {xu2023llmcad}
\bibfield{author}{\bibinfo{person}{Daliang Xu}, \bibinfo{person}{Wangsong Yin}, \bibinfo{person}{Xin Jin}, \bibinfo{person}{Ying Zhang}, \bibinfo{person}{Shiyun Wei}, \bibinfo{person}{Mengwei Xu}, {and} \bibinfo{person}{Xuanzhe Liu}.} \bibinfo{year}{2023}\natexlab{}.
\newblock \showarticletitle{Llmcad: Fast and scalable on-device large language model inference}.
\newblock \bibinfo{journal}{\emph{arXiv preprint arXiv:2309.04255}} (\bibinfo{year}{2023}).
\newblock


\bibitem[Xu et~al\mbox{.}(2022)]%
        {xu2022systematic}
\bibfield{author}{\bibinfo{person}{Frank~F Xu}, \bibinfo{person}{Uri Alon}, \bibinfo{person}{Graham Neubig}, {and} \bibinfo{person}{Vincent~Josua Hellendoorn}.} \bibinfo{year}{2022}\natexlab{}.
\newblock \showarticletitle{A systematic evaluation of large language models of code}. In \bibinfo{booktitle}{\emph{ACM MAPS}}.
\newblock


\bibitem[Xu et~al\mbox{.}(2024)]%
        {xu2024penetrative}
\bibfield{author}{\bibinfo{person}{Huatao Xu}, \bibinfo{person}{Liying Han}, \bibinfo{person}{Qirui Yang}, \bibinfo{person}{Mo Li}, {and} \bibinfo{person}{Mani Srivastava}.} \bibinfo{year}{2024}\natexlab{}.
\newblock \showarticletitle{Penetrative ai: Making llms comprehend the physical world}. In \bibinfo{booktitle}{\emph{HotMobile}}. \bibinfo{pages}{1--7}.
\newblock


\bibitem[Yang et~al\mbox{.}(2024)]%
        {yang2024drhouse}
\bibfield{author}{\bibinfo{person}{Bufang Yang}, \bibinfo{person}{Siyang Jiang}, \bibinfo{person}{Lilin Xu}, \bibinfo{person}{Kaiwei Liu}, \bibinfo{person}{Hai Li}, \bibinfo{person}{Guoliang Xing}, \bibinfo{person}{Hongkai Chen}, \bibinfo{person}{Xiaofan Jiang}, {and} \bibinfo{person}{Zhenyu Yan}.} \bibinfo{year}{2024}\natexlab{}.
\newblock \showarticletitle{DrHouse: An LLM-empowered Diagnostic Reasoning System through Harnessing Outcomes from Sensor Data and Expert Knowledge}.
\newblock \bibinfo{journal}{\emph{arXiv preprint arXiv:2405.12541}} (\bibinfo{year}{2024}).
\newblock


\bibitem[Yao et~al\mbox{.}(2023)]%
        {yao2023react}
\bibfield{author}{\bibinfo{person}{Shunyu Yao}, \bibinfo{person}{Jeffrey Zhao}, \bibinfo{person}{Dian Yu}, \bibinfo{person}{Nan Du}, \bibinfo{person}{Izhak Shafran}, \bibinfo{person}{Karthik Narasimhan}, {and} \bibinfo{person}{Yuan Cao}.} \bibinfo{year}{2023}\natexlab{}.
\newblock \showarticletitle{ReAct: Synergizing Reasoning and Acting in Language Models}. In \bibinfo{booktitle}{\emph{ICLR}}.
\newblock


\bibitem[Yi et~al\mbox{.}(2023)]%
        {yi2023edgemoe}
\bibfield{author}{\bibinfo{person}{Rongjie Yi}, \bibinfo{person}{Liwei Guo}, \bibinfo{person}{Shiyun Wei}, \bibinfo{person}{Ao Zhou}, \bibinfo{person}{Shangguang Wang}, {and} \bibinfo{person}{Mengwei Xu}.} \bibinfo{year}{2023}\natexlab{}.
\newblock \showarticletitle{Edgemoe: Fast on-device inference of moe-based large language models}.
\newblock \bibinfo{journal}{\emph{arXiv preprint arXiv:2308.14352}} (\bibinfo{year}{2023}).
\newblock


\bibitem[Yin et~al\mbox{.}(2024)]%
        {yin2024llm}
\bibfield{author}{\bibinfo{person}{Wangsong Yin}, \bibinfo{person}{Mengwei Xu}, \bibinfo{person}{Yuanchun Li}, {and} \bibinfo{person}{Xuanzhe Liu}.} \bibinfo{year}{2024}\natexlab{}.
\newblock \showarticletitle{Llm as a system service on mobile devices}.
\newblock \bibinfo{journal}{\emph{arXiv preprint arXiv:2403.11805}} (\bibinfo{year}{2024}).
\newblock


\bibitem[Zhang et~al\mbox{.}(2024)]%
        {zhang2024llamatouch}
\bibfield{author}{\bibinfo{person}{Li Zhang}, \bibinfo{person}{Shihe Wang}, \bibinfo{person}{Xianqing Jia}, \bibinfo{person}{Zhihan Zheng}, \bibinfo{person}{Yunhe Yan}, \bibinfo{person}{Longxi Gao}, \bibinfo{person}{Yuanchun Li}, {and} \bibinfo{person}{Mengwei Xu}.} \bibinfo{year}{2024}\natexlab{}.
\newblock \showarticletitle{LlamaTouch: A Faithful and Scalable Testbed for Mobile UI Task Automation}.
\newblock \bibinfo{journal}{\emph{ACM UIST}} (\bibinfo{year}{2024}).
\newblock


\bibitem[Zhong et~al\mbox{.}(2024)]%
        {zhong2024ldb}
\bibfield{author}{\bibinfo{person}{Li Zhong}, \bibinfo{person}{Zilong Wang}, {and} \bibinfo{person}{Jingbo Shang}.} \bibinfo{year}{2024}\natexlab{}.
\newblock \showarticletitle{LDB: A Large Language Model Debugger via Verifying Runtime Execution Step-by-step}. In \bibinfo{booktitle}{\emph{ACL: Findings.}}
\newblock


\bibitem[Ziegler et~al\mbox{.}(2022)]%
        {ziegler2022productivity}
\bibfield{author}{\bibinfo{person}{Albert Ziegler}, \bibinfo{person}{Eirini Kalliamvakou}, \bibinfo{person}{X~Alice Li}, \bibinfo{person}{Andrew Rice}, \bibinfo{person}{Devon Rifkin}, \bibinfo{person}{Shawn Simister}, \bibinfo{person}{Ganesh Sittampalam}, {and} \bibinfo{person}{Edward Aftandilian}.} \bibinfo{year}{2022}\natexlab{}.
\newblock \showarticletitle{Productivity assessment of neural code completion}. In \bibinfo{booktitle}{\emph{ACM MAPS}}.
\newblock


\end{thebibliography}
\end{document}